\documentclass[prd,superscriptaddress,twocolumn,floatfix,amsmath,amssymb,amsfonts,nofootinbib,longbibliography]{revtex4-2}

\usepackage{float} 
\usepackage{ragged2e}
\usepackage{scalerel}
\usepackage[normalem]{ulem}
\usepackage[english]{babel}
\usepackage{graphicx}
\usepackage{dcolumn}
\usepackage{bm}
\usepackage{blindtext}
\usepackage{verbatim}
\usepackage{relsize}
\usepackage{mathrsfs}
\usepackage{musicography}
\usepackage{amsmath}
\usepackage{blindtext}
\usepackage{cancel}
\usepackage{physics}
\usepackage{epstopdf}
\usepackage{mathtools}
\usepackage{blindtext}
\usepackage{tensor}
\usepackage{color}
\usepackage[usenames,dvipsnames]{pstricks}
\usepackage{epsfig}
\usepackage{pst-grad} 
\usepackage{pst-plot} 
\usepackage{hyperref}
\usepackage{verbatim}
\usepackage{slashed}
\usepackage{caption}
\usepackage{subcaption}
\usepackage{dsfont}
\usepackage[english]{babel}
\usepackage{amsmath,amssymb,amsthm}  
\usepackage{todonotes}

\newcommand{\mf}{\mathsf}

\newcommand{\ii}{\mathrm{i}}
\newcommand{\RR}{\mathbb{R}}
\newcommand{\ee}{\mathrm{e}}

\newcommand{\s}{\hat{\sigma}}
\newcommand{\tc}[1]{\textsc{#1}}

\newcommand{\F}{\mathcal{F}}

\allowdisplaybreaks[1] 

\begin{document}

\title{Ambient temperature versus ambient acceleration in the circular motion Unruh effect
}

\author{Cameron R. D. Bunney}
\email{cameron.bunney@nottingham.ac.uk}

\affiliation{School of Mathematical Sciences, University of Nottingham, Nottingham NG7 2RD, UK}

\author{Leo Parry}
\email{leo.parry@nottingham.ac.uk}

\affiliation{School of Mathematical Sciences, University of Nottingham, Nottingham NG7 2RD, UK}

\author{T. Rick Perche}
\email{trickperche@perimeterinstitute.ca}

\affiliation{Department of Applied Mathematics, University of Waterloo, Waterloo, Ontario, N2L 3G1, Canada}
\affiliation{Institute for Quantum Computing, University of Waterloo, Waterloo, Ontario, N2L 3G1, Canada}
\affiliation{Perimeter Institute for Theoretical Physics, Waterloo, Ontario, N2L 2Y5, Canada}

\author{Jorma Louko}
\email{jorma.louko@nottingham.ac.uk}

\affiliation{School of Mathematical Sciences, University of Nottingham, Nottingham NG7 2RD, UK}

\date{October 2023, revised February 2024. \\ aaPublished in Phys.\ Rev.\ D \textbf{109}, 065001 (2024), doi.org/10.1103/PhysRevD.109.065001.\\ aaFor Open Access purposes, this Author Accepted Manuscript is made available under CC BY public copyright.}

\begin{abstract}
It is well known that the experience of a linearly accelerated observer with acceleration $a$, interacting with a massless scalar field in its vacuum state in $3+1$ Minkowski spacetime, is identical to that of a static observer interacting with a massless scalar field in a thermal state of temperature $a/2\pi$ in $3+1$ Minkowski spacetime. We study the robustness of this duality by comparing an observer undergoing circular motion in a thermal bath with an observer that undergoes circular motion around a linearly accelerated trajectory. We find that in most regimes, observers in these two cases experience the field in different ways, and are generally able to tell the difference between the two cases by measuring observables localized along their trajectories.
\end{abstract}

\maketitle

\section{Introduction}
{In investigating black hole evaporation Unruh considered a uniformly accelerated observer in Minkowski spacetime probing a massless scalar field in its vacuum state.} The result of this thought experiment is now known as the Unruh effect and states that the uniformly linearly accelerated observer would experience a thermal bath of particles with temperature given by~\cite{Unruh1976}
\begin{equation}\label{eq:TU}
    T_U ~=~ \frac{\hbar a}{2\pi c k_B}\,,
\end{equation}
where $a$ is the proper acceleration of the trajectory, $\hbar$ is Planck's constant, $c$ is the speed of light, $k_B$ is the Boltzmann constant, and $T_U$ is referred to as the Unruh temperature. In essence, the Unruh effect states that there is a duality between a static observer probing a thermal field of temperature $a/2\pi$ and a uniformly linearly accelerated observer with proper acceleration $a$ probing the Minkowski vacuum. This duality becomes even more precise in the case where the observer probes a massless scalar field in $3+1$-dimensional Minkowski spacetime, in which case the field correlations are \emph{identical} in the two cases.

Although the Unruh effect is widely accepted within quantum field theory, there has been no direct experimental verification of the phenomenon yet. The main reason for this is that an acceleration of $10^{20}~\mathrm{ms}^{-2}$ would be required to register an Unruh temperature of $1~\mathrm{K}$. There are, however, experimental proposals for detecting an analogue of the Unruh effect within analogue gravity which utilise condensed matter systems~\cite{unruhExp,bunnySounds}, where the speed of light is replaced by the speed of sound in the medium. For example, in superfluid helium, working with a thin film of $100~\mathrm{nm}$, the acceleration required for a $1~\mathrm{K}$ analogue Unruh temperature is reduced by $10$ orders of magnitude~\cite{bunnySounds}. In addition, the experimental proposals use a circular trajectory, which has long-standing experimental and theoretical interest~\cite{BellUnruh,Unruhelectron,Biermann:2020bjh}. The circular trajectory enjoys numerous advantages over linear acceleration; in particular, it is a bounded motion, allowing for arbitrary interaction times within a finite laboratory.

Unlike in uniform linear acceleration, observers undergoing uniform circular motion do not experience a true thermal bath when interacting with a quantum field in its vacuum state. The observers' experiences can, however, be described in terms of an effective energy-dependent temperature by fitting the excitation and de-excitation rates within a limited energy interval to Einstein's detailed balance formula~\cite{Einstein, terhaar-book}. We refer to this notation of temperature as the detailed balance temperature~\cite{Unruhelectron, Biermann:2020bjh}.

When attempting to probe the Unruh effect, it is important to bear in mind that all laboratories operate at nonzero temperatures. Indeed, the lowest-measured temperatures in Bose-Einstein condensates (BECs) were of the order of picokelvin~\cite{BECTemp1,BECTemp2}. For this reason, when one considers the circular motion Unruh effect, it is relevant to consider the robustness of the effect to a background thermal bath. Moreover, in the case of a massless scalar field in $3+1$ spacetime dimensions, there is a duality between observers with acceleration $a$ probing the vacuum and static observers probing a state of temperature $T_U$. 
It is therefore natural to ask the following:
\begin{center}
    
\textit{Does an observer in circular motion through a thermal bath of temperature $T_U$ respond in the same way as an observer undergoing uniform acceleration as well as circular motion in the plane orthogonal to its acceleration?}
\end{center}
 Answering this question is the primary goal of this paper.

In addressing this, we make use of particle detectors coupled to a massless scalar field. The concept of a detector used to operationally probe the particle content of a quantum field was introduced by Unruh~\cite{Unruh1976} and later refined by DeWitt~\cite{DeWitt}. Such particle detector models are referred to as Unruh-DeWitt (UDW) detectors. A UDW detector is a localized quantum system which interacts with a quantum field and thus acts as a measurement device, quantifying the excitations and de-excitations of the field. In this way, the device could also be used as a thermometer to probe the temperature of the field. This simple tool has also proven to be ideal for probing quantum fields with ubiquitous applications within quantum field theory (QFT), including: measuring Hawking radiation~\cite{Unruh1976,bhDetectors} and the Unruh effect~\cite{Unruh1976,Matsas1,Unruh-Wald,Takagi,matsasUnruh,antiUnruh,unruhSlow,mine}; measuring entanglement in quantum field theory~\cite{Valentini1991,Reznik2003,Pozas-Kerstjens:2015,Pozas2016,ericksonNew,hectorMass}; accessing the correlation function of quantum fields~\cite{pipo,geometry}; and the implementation of numerous quantum information protocols~\cite{teleportation,Jonsson1,Jonsson2,Jonsson3,Jonsson4}. Moreover, there are many regimes where particle detectors can be used to model physical probes which can be implemented in a laboratory, such as atoms interacting with the electromagnetic field~\cite{Pozas2016,Nicho1,richard}, nucleons interacting with the neutrino field~\cite{neutrinos,carol}, among others~\cite{pitelli,boris,generalPD}. In particular, the UDW model can also be used to describe the proposals of measuring the circular Unruh effect~\cite{unruhExp,bunnySounds}.

By considering particle detectors coupled to a massless scalar field in $3+1$ spacetime dimensions, we compare the case where the detector undergoes uniform acceleration together with circular motion interacting with the vacuum and the case where the detector undergoes circular motion and interacts with a thermal field at the Unruh temperature. We explore different regimes of both cases, and compare them in Section~\ref{sec:versus}. Overall, we find that any kind of circular motion around the trajectory is in principle enough to distinguish the uniform acceleration from a thermal bath, as the temperature experienced in each case always differs. Nevertheless, we find regimes such that the two cases become almost indistinguishable, such as for a small circular trajectory radius, and for angular velocities smaller than the energy gap of the detector. Overall, through analytical techniques we study the transition rate of the detectors, and present a general comparison of the two cases in many limiting regimes.

The paper is organized as follows. In Section~\ref{sec:UDW}, we review the UDW model, presenting the well-known duality between uniform acceleration and a static detector in a thermal bath. In Section~\ref{sec:hypertor} we consider a UDW detector undergoing circular motion and uniform acceleration in an orthogonal direction. In Section~\ref{sec:centerOfMass} we consider a detector undergoing circular motion while in a thermal bath. In Section~\ref{sec:versus} we compare the two cases, and show under which conditions they are distinguishable. The conclusions of our work can be found in Section~\ref{sec:conclusions}. Other than in Eq.~\eqref{eq:TU}, we use units where $c = \hbar = k_B = 1$. In asymptotic formulae, $f(x)=\mathcal{O}\left(g(x)\right)$ denotes that $f(x)/g(x)$ remains bounded in the limit considered, $f(x)=o\left(g(x)\right)$ denotes that $f(x)/g(x)$ tends to zero in the limit considered, and $f(x)\sim g(x)$ denotes that $f(x)/g(x)$ tends to one in the limit considered.

\section{UDW detector model and temperature}\label{sec:UDW}

In this section, we review the two-level pointlike Unruh-DeWitt (UDW) particle detector model. This model consists of a qubit undergoing a timelike trajectory $\mf z(\tau)$ in a background spacetime, which we assume to be $3+1$ Minkowski spacetime. 
{The Hilbert space of the detector $\mathcal{H}_\tc{d}\simeq\mathbb{C}^2$ is spanned by the orthonormal basis $\{\ket{0},\ket{1}\}$. The internal dynamics of the qubit are described by a Hamiltonian $\hat{H}_{\mathrm{D}}$ whose action on $\mathcal{H}_{\mathrm{D}}$ is $H_{\tc{d}}\ket{0}=0$ and $\hat{H}_{\tc{d}}\ket{1}=E\ket{1}$. This describes a two-level system with an energy gap $|E|$. For $E>0$, $\ket{0}$ is the ground state and $\ket{1}$ the excited state. For $E<0$, the roles are reversed.}

We consider the case where the particle detector is coupled to a massless, real scalar quantum field $\hat{\phi}(\mf x)$. In the Hilbert space representation associated to the translation-invariant Minkowski vacuum $\ket{0}$, the field can be written in terms of plane-wave modes $u_{\bm k}(\mf x)$ as
\begin{align}
    \hat{\phi}(\mf x) &~=~  \int_{\RR^3}\dd^3 \bm k\left( u_{\bm k}(\mf x) \hat{a}_{\bm k}+u_{\bm k}^*(\mf x) \hat{a}^\dagger_{\bm k} \right)\,,\\ 
    u_{\bm k}(\mf x) &~=~ \frac{1}{(2\pi)^{\frac{3}{2}}}\frac{\ee^{\ii \mf k \cdot \mf x}}{\sqrt{2|\bm k|}}\,,
\end{align}
where $\mf k = (|\bm k|, \bm k)$, $\mf x = (t,\bm x)$ is an inertial coordinate system, and $\mf k \cdot \mf x = \eta_{\mu\nu}k^\mu x^\nu$, with the Minkowski metric $\eta_{\mu\nu} = \text{diag}(-1,1,1,1)$. The operators $\hat{a}_{\bm k}$ and $\hat{a}^\dagger_{\bm k}$ are the annihilation and creation operators, which satisfy the canonical commutation relations
\begin{equation}
    \comm{\hat{a}_{\bm k}}{\hat{a}^\dagger_{\bm k'}} ~=~ \delta^{(3)}(\bm k - \bm k')\,.
\end{equation}
The Hilbert space associated to this representation can then be constructed by repeatedly acting on the vacuum state $\ket{0}$ with the creation operators $\hat{a}^\dagger_{\bm k}$. 

The interaction between the qubit and the field is described by a linear coupling of the detector's monopole moment $\hat{\mu} = \s^+ + \s^-$ with the field amplitude $\hat{\phi}(\mf x)$. Here $\s^+$ and $\s^-$ are the raising and lowering operators in $\mathcal{H}_\tc{d}$. In the interaction picture, the interaction Hamiltonian can be written as
\begin{equation}\label{eq:HI}
    \hat{H}_I(\tau) ~=~ \lambda \chi(\tau) \hat{\mu}(\tau) \hat{\phi}(\mf z(\tau))\,,
\end{equation}
where $\lambda$ is a dimensionless coupling constant and $\chi(\tau)$ is the switching function, which specifies how the interaction is turned on and off. The time-evolved monopole moment $\hat{\mu}(\tau)$ in the interaction picture is given by
\begin{equation}
    \hat{\mu}(\tau) ~=~ \ee^{\ii E \tau} \s^+ + \ee^{-\ii E \tau} \s^-\,.
\end{equation}
With this simple formulation, one can model absorption and emission by the detector within a time frame defined by the support of the switching function $\chi(\tau)$. For instance, if a detector starts in its ground state and transitions to its excited state after interaction with the quantum field, we say that the detector has \textit{clicked} and registered a particle. It is in this sense that this model is referred to as a particle detector.

A relevant quantity that can be used to characterize the interaction of a particle detector with the field is its \emph{leading-order} transition probability, once one traces over the field degrees of freedom. Considering the transition from $\ket{0}$ to $\ket{1}$, the sign of $E$ specifies whether this is an excitation ($E>0$) or de-excitation ($E<0$). This transition probability is given by
\begin{equation}\label{eq:PE}
    P(E) = \lambda^2 \int_{\RR^2 } \dd \tau \dd \tau' \chi(\tau) \chi(\tau') \ee^{- \ii E(\tau - \tau')}\mf W_{\hat{\rho}}(\mf z(\tau), \mf z(\tau'))\,,
\end{equation}
where $\mf W_{\hat{\rho}}(\mf x, \mf x') = \text{tr}(\hat{\rho}\, \hat{\phi}(\mf x) \hat{\phi}(\mf x'))$ is the two-point function (or Wightman function) of the field in the state $\hat{\rho}$. It should be understood as the integral kernel of a bidistribution. Explicitly, for a massless field in the Minkowski vacuum, we have
\begin{align}
    \mf W_0(\mf x, \mf x') &~=~ \bra{0}\!\hat{\phi}(\mf x) \hat{\phi}(\mf x') \!\ket{0} ~=~\frac{1}{4\pi^2 \sigma_\varepsilon(\mf x, \mf x')}\,,\\
    \sigma_\varepsilon(\mf x, \mf x') &~=~ - (t-t' - \ii \varepsilon)^2 + (\bm x - \bm x')^2\,,
\end{align}
where the limit $\varepsilon \to 0^+$ is taken after integration. As a simple example, consider an inertial detector with a Gaussian switching function of the form
\begin{equation}\label{gaussian switching}
    \chi_\tc{g}(\tau) ~=~ \ee^{- \pi \tau^2/2T^2}\,.
\end{equation}
Note that this expression is conveniently normalized (see Appendix \ref{app:rates}) and $T$ can be used to control the time of the switching of the interaction. In this case, one can show~\cite{hectorMass} that the leading-order transition probability of the detector is given by
\begin{equation}
    P_\tc{g}(E) ~=~ \frac{\lambda^2 }{4 \pi}\left(\ee^{-E^2T^2/\pi}-E T \operatorname{erfc}\left(\frac{E T}{\sqrt{\pi}}\right)\right)\,,
\end{equation}
where $\text{erfc}$ denotes the complementary error function~\cite{database}. 

From~\eqref{eq:PE}, it is clear that the transition probabilities of the detector depend on the specific shape of the switching function $\chi(\tau)$. This is because the interaction Hamiltonian~\eqref{eq:HI} is explicitly time dependent due to the switching on and off of the interaction of the detector with the field. One way of obtaining a result that is independent of the switching is to consider the case where the interaction is always turned on (that is, $\chi(\tau) = 1$). This case can be obtained by adiabatically scaling the switching function: keeping its maximum value constant, but pushing the switch-on and switch-off to the infinitely far past and future, respectively. 

However, in this case, perturbation theory fails to hold because the integral of Eq.~\eqref{eq:PE} grows linearly with the time duration of the interaction. The way to handle this issue is to consider the detector's transition rate $\F(E)$, which can be defined as
\begin{equation}
    \F(E) ~=~ \lim_{T\rightarrow \infty} \frac{1}{\lambda^2}\frac{P(E)}{T}\,,
\end{equation}
where $T$ is a suitable time parameter that controls the time duration of the interaction and we have divided by the coupling constant for convenience. Intuitively, the transition rate can be thought of as 
the number of transitions of an ensemble of detector per unit time~\cite{LoukoCurvedSpacetimes}. For instance, using the Gaussian switching of Eq.~\eqref{gaussian switching}, we can compute the transition rate
\begin{equation}
    \F(E) ~=~ \lim_{T\rightarrow \infty} \frac{P_\tc{g}(E)}{T} ~=~ -\frac{E}{2\pi}\Theta(-E)\,,
\end{equation}
where $\Theta(x)$ denotes the Heaviside theta function~\cite{database}. The result above can be easily obtained using the fact that $\lim_{a\to\infty}\operatorname{erfc}(a x) = 2 \Theta(-x)$. Hence, if a detector starts in the ground state ($E>0$) while interacting with the Minkowski vacuum, it will not emit any particles. On the other hand, if the detector starts in the excited state ($E<0$), it will emit particles at a rate proportional to its energy gap $E$.

It must be mentioned that there is nothing particular in choosing a Gaussian switching function to obtain the limit of the transition rate of the detector, and any other switching function might have been chosen for this calculation with a suitable rescaling of the time parameter $T$. For more details regarding this discussion, we refer the reader to Appendix~\ref{app:rates}. In this paper, the Gaussian transition probability will be a convenient choice when we study the response of a detector in the small-gap limit.

A particular class of trajectories amenable to analytic exploration is that of stationary trajectories, the integral curves of Killing vector fields. If the field is prepared in a stationary state, the field's Wightman function can be written as a function of the difference in proper time along the trajectory,
\begin{equation}\label{stationary wightman}
    \mf W(\mf z(\tau),\mf z(\tau')) ~=~ \mf W(\mf z(\tau - \tau'), \mf z(0)) ~\equiv~ W(\tau - \tau')\,.
\end{equation}
That is, the pullback of the Wightman function to the detector's world line can be written as the distribution $W(\tau - \tau')$. A full classification of the stationary trajectories in Minkowski spacetime, with the field in the Minkowski vacuum, was analyzed in~\cite{good, Bunneystationary}. In these cases, it is possible to rewrite the detector's transition rate as a single integral. In order to see this, we consider again the switching function of Eq.~\eqref{gaussian switching}, and assume the detector's trajectory and the field's initial state to be stationary. In this case, we can write the transition probability as
\begin{multline}
    P(E) ~=~ \lambda^2 \int_{\RR^2 }\dd \tau \dd \tau'\, \ee^{- \pi \tau^2/2 T^2}\ee^{- \pi \tau'{}^2/2 T^2} \\\times \ee^{-\ii E(\tau - \tau')}W(\tau - \tau')\,.
\end{multline}
We then perform a change of variables $u = \tau - \tau'$ and $v = (\tau + \tau')/2$, such that $\tau^2 + \tau'{}^2  = u^2/2 + 2v^2$. This decouples the integrals over $u$ and $v$. Performing the integral over $v$ gives a factor of $T$, so that we are left with a single integral over $u$,
\begin{align}
    P(E) &~=~ \lambda^2 \int_{\RR^2 } \dd u \dd v\, \ee^{- \pi u^2/4 T^2}\, \ee^{- \pi {v}^2/T^2}\ee^{-\ii Eu} W(u)\,,\nonumber\\ &~=~ \lambda^2 T \int_{\RR} \dd u\,  \ee^{- \pi u^2/4 T^2} \ee^{-\ii Eu}W(u)\,.
\end{align}
After relabelling $u$ to $\tau$, one can express the transition rate as the limit of a single integral which corresponds to the Fourier transform of the function $W(\tau)$,
\begin{align}
    \F(E) &~=~ \lim_{T \to \infty}\int_\RR \dd \tau \, \ee^{- \pi \tau^2/4 T^2}\ee^{- \ii E \tau} W(\tau)\,,\nonumber\\ &~=~ \int_\RR \dd \tau \, \ee^{- \ii E \tau} W(\tau)\,.\label{eqn: F def}
\end{align}
{We remind the reader that the notation in~\eqref{eqn: F def} suppresses the $-\ii\varepsilon$ regularization that defines the Wightman function.} In the case of an inertial detector interacting with the Minkowski vacuum, this integral reads~\cite{Biermann:2020bjh}
\begin{equation}\label{minkowski response}
    \F_0(E) ~=~ - \frac{E}{2\pi}\Theta(-E)\,.
\end{equation}
We stress that this result is not particular to the specific choice of Gaussian switching, as may be shown by the techniques of~\cite{waitUnruh}, and we again refer the reader to the general argument in Appendix~\ref{app:rates}.

A noteworthy example is that of a uniformly linearly accelerated detector with constant proper acceleration $a$, from which originates the Unruh effect~\cite{Unruh1976}. This trajectory is the orbit of a boost Killing vector of Minkowski spacetime, and is thus  stationary. In this case, the pullback of the Wightman function to the trajectory, $W_a(\tau)$, can be written as
\begin{equation}\label{accel Wightman}
    W_a(\tau) ~=~ - \frac{1}{4\pi^2}\frac{1}{\frac{4}{a^2}\sinh^2(\tfrac{a}{2}(\tau-\ii\varepsilon))}\,.
\end{equation}
The response function is given by
\begin{equation}\label{eqn:lin accel}
    \mathcal{F}_a(E) ~=~ \frac{E}{2\pi}\frac{1}{\ee^{2\pi E/a} - 1}\,,
\end{equation}
which satisfies
\begin{equation}
     \frac{\F_a(-E)}{\F_a(E)} ~=~ \ee^{2\pi E/a}\,, 
\label{eq:Fa-detailedbalance}
\end{equation} 
which is Einstein's detailed balance condition \cite{Einstein} in the Unruh temperature $T_U = a/(2\pi)$.

The case of a uniformly linearly accelerated detector is indeed completely dual to the case of a static detector in the Minkowski vacuum interacting with a thermal bath of temperature $a/(2\pi)$: if the field is initially at a Kubo-Martin-Schwinger (KMS) state of inverse temperature $\beta$ with respect to the inertial notion of time translation associated with the detector's motion, the pullback of the Wightman function can be written as \cite{Takagi}
\begin{equation}\label{thermal bath wightman}
    W_\beta(\tau) ~=~ -\frac{1}{4 \beta^2\sinh^2(\tfrac{\pi}{\beta}(\tau-\ii\varepsilon))}\,,
\end{equation}
which is identical to \eqref{accel Wightman} for $\beta=2\pi/a$. This firmly establishes the duality between a detector in linearly accelerated motion with acceleration $a$ in the vacuum and a static detector in a thermal bath of temperature $a/(2\pi)$, for a massless quantum field in $3+1$ spacetime dimensions.

Three comments are in order.

First, when the field has mass, or when the spacetime dimension differs from $(3+1)$, or when the field has a nonzero spin, or when the coupling to the field is nonlinear, the vacuum response functions in uniform linear acceleration and the inertial response function in a static thermal bath need no longer be identical, but the uniform linear acceleration response function still satisfies the KMS property \eqref{eq:Fa-detailedbalance} \cite{Takagi}, as follows from the KMS property of the Minkowski vacuum when written in terms of excitations on the Rindler vacuum. This is a consequence of the Bisognano-Wichmann theorem, which characterises the behaviour of the Minkowski-vacuum correlation functions under boosts~\cite{Bisognano1975,Bisognano1976}.

Second, while the above considerations hold for the transition rates, they imply that under the Markovian approximation the detector's asymptotic final state is also thermal. In a slightly generalised setting, suppose that the transition rate of a detector in stationary motion satisfies a generalised version of Einstein's detailed balance, 
\begin{equation}\label{detailed balance}
    \frac{\F(-E)}{\F(E)} ~=~ \ee^{\beta(E) E}\,,
\end{equation}
where $\beta(E)$ is interpreted as a generalised, energy-dependent inverse temperature.  Then, in the Markovian approximation, the detector's asymptotic final state is \cite{assymtoticBenito} 
\begin{equation}\label{gives final state, Mr}
    \hat{\rho}_\tc{d} ~=~ \frac{1}{1 + \ee^{- \beta(E)E}}\begin{pmatrix}
        1 & 0 \\
        0 & \ee^{- \beta(E)E}
    \end{pmatrix}\,.
\end{equation}
If $\beta(E)$ is not constant, this final state can be interpreted as a nonequilibrium state of effective energy-dependent temperature $1/\beta(E)$.

Third, while the Unruh temperature in uniform linear acceleration is universal, independent of the details of the field, the energy-dependent inverse temperature $\beta(E)$ in a general stationary motion, defined by \eqref{detailed balance}, may depend also on the characteristics of the field. From now on we specialise to a massless scalar field in 3+1 spacetime dimensions.

A trajectory of both theoretical and experimental interest is that of uniform circular motion. Recent experimental proposals~\cite{unruhExp, bunnySounds} suggest using this motion for the experimental realization of an analogue of the Unruh effect, due to its greater experimental implementability. In particular, circular motion remains bounded within a finite size for an arbitrarily long interaction time. 
It is well documented, however, that circular motion does not produce a truly thermal response~\cite{BellUnruh, Biermann:2020bjh}, though the effective temperature may be approximated as thermal throughout most of its parameter space. Furthermore, investigations into the inclusion of an initial thermal state~\cite{bunnyCircular} reveal that acceleration dependence still remains.

In this paper, we consider a related question: how similar are the thermal baths provided by the Unruh effect and an initial thermal state for a detector in uniform circular motion? That is, we consider circular motion in a $3+1$ thermal bath at rest with respect to the trajectory and compare it to a detector in circular motion in the $xy$-plane but accelerated parallel to the $z$-axis, which has been termed hypertor~\cite{good} motion.

\section{Hypertor Motion}\label{sec:hypertor}

In this section, we describe a detector undergoing hypertor motion in Minkowski spacetime~\cite{good}. This motion consists of uniform acceleration in one direction combined with circular motion in the plane orthogonal to it. Without loss of generality, we assume the acceleration to be along the $z$-axis and the circular motion to be within the $xy$-axis in inertial coordinates. With these choices the detector's trajectory is parametrized by proper time $\tau$ as

\begin{align}\nonumber
    \mf z (\tau) &~=~ \left(\frac{1}{a}\sinh(a\gamma \tau), R \cos(\Omega \gamma \tau),\right. \\&\left.\quad\quad\quad\quad\quad\quad R \sin(\Omega \gamma \tau),\frac{1}{a}\cosh(a\gamma\tau)\right)\,,\\
    \gamma&~=~\frac{1}{\sqrt{1-R^2\Omega^2}}\,,\label{eqn: gamma def}
\end{align}
where $a>0$ is the proper acceleration of the center of the circular motion, and $R>0$ and $\Omega>0$ are the radius and angular velocity of the circular motion, as measured in the frame of the uniformly accelerated center, with $R\Omega<1$. $\gamma$ can be interpreted as a redshift factor between the circular motion and the uniformly accelerated trajectory in its center.

Note that in order to ensure that the trajectory is  timelike, we must have $R\Omega < 1$. This trajectory is the composition of uniformly linearly accelerated motion and uniform circular motion; as such, it is the orbit of the combination of a boost Killing vector and a rotational Killing vector. Hence, the trajectory is stationary.

The pullback of the Wightman function along the trajectory can be written as
\begin{multline}\label{wightman hypertor}
    W_{\tc{h}}( \tau) ~=~ \\- \frac{1}{4\pi^2} \frac{1}{\tfrac{4}{a^2}\sinh^2(\frac{\gamma a}{2}(\tau-\ii\varepsilon)) - 4R^2 \sin^2(\frac{\gamma \Omega}{2}(\tau-\ii\varepsilon))}\,,
\end{multline}
where we use the subscript $\tc{H}$ to label the hypertor motion. The detector's transition rate can then be written as
\begin{align}\label{hypertor response}\nonumber
    \F_\tc{h}(E) &~=~ \int_\RR \dd \tau\, \ee^{- \ii E\tau}W_{\tc{h}}(\tau)\,,\\\nonumber&~=~ - \frac{1}{4\pi^2}\int_\RR \dd \tau\,\ee^{-\ii E\tau} \\\times&\frac{1}{\tfrac{4}{a^2}\sinh^2(\frac{\gamma a}{2}(\tau-\ii\varepsilon)) - 4R^2 \sin^2(\frac{\gamma \Omega}{2}(\tau-\ii\varepsilon))}\,.
\end{align}As we have seen in Eq.~\eqref{gives final state, Mr}, the integral above yields the detector's final state and its effective temperature after interaction with the field for a sufficiently long time~\cite{waitUnruh}. However, the integral in Eq.~\eqref{hypertor response} cannot be evaluated in terms of known functions. Therefore, we must analyze it in limiting regimes to get an analytic grasp of the detector's behaviour.

\subsection{Small-gap limit}\label{subsec: small gap HT}
We consider first the limit of a small energy gap, $E\rightarrow 0$ with all other parameters fixed. We find a power series expansion in $E$ to all orders which displays nontrivial behaviour. To perform a small-$E$ expansion, we follow the method detailed in~\cite{Biermann:2020bjh, handsome}, which regulates the integral by adding and subtracting the singular vacuum term
\begin{equation}\label{hadamard ht}
    W_{\tc{h}}(\tau) ~\sim~  -\frac{1}{4\pi^2 (\tau-\ii\varepsilon)^2}\,.
\end{equation}
This results in
\begin{widetext}
\begin{equation}\label{eqn: Fsplit}
    \F_\tc{h}(E) ~=~ \underbrace{- \frac{1}{4\pi^2}\int_\RR \dd \tau\, \cos(E \tau)\left(\frac{1}{\tfrac{4}{a^2}\sinh^2(\frac{\gamma a \tau}{2}) - 4R^2 \sin^2(\frac{\gamma \Omega \tau}{2})} - \frac{1}{\tau^2}\right)}_{\textbf{(a)}} -  \underbrace{\frac{1}{4\pi^2}\int_\RR\dd\tau\, \ee^{-\ii E\tau}\frac{1}{(\tau-\ii\varepsilon)^2}}_{\textbf{(b)}}\,.
\end{equation}
\end{widetext}
By grouping the addition and subtraction as in~\eqref{eqn: Fsplit}, the integral splits into a regular part \textbf{(a)} and a distributional contribution \textbf{(b)}. One should note that the small-$\tau$ behaviour in \textbf{(a)} is divergence free, hence this integral is well-defined and the limit $\varepsilon\rightarrow0^+$ can be taken. Hence, integral \textbf{(a)} is even in $\tau$, which allows us to replace $\exp(\ii E\tau)$ by $\cos(E\tau)$. Integral \textbf{(b)} still requires the regulating $-\ii\varepsilon$ and corresponds to the distributional behaviour of the Wightman function. In particular, it is the inertial transition rate in Minkowski spacetime. An application of the residue theorem on integral \textbf{(b)} results in $\F_0(E)$, given by Eq.~\eqref{minkowski response}. This allows us to write
\begin{equation}
    \F_\tc{h}(E) ~=~ \F_0(E)+\F_{\tc{h}}^{\text{corr}}(E)\,,
\end{equation}
where $\mathcal{F}_0(E)$ is given in~\eqref{minkowski response} and
\begin{multline}\label{eq:Fcorr}
    \F_{\tc{h}}^\text{corr}(E) ~=~ - \frac{1}{4\pi^2}\int_\RR \dd \tau\, \cos(E \tau)\\\times\left(\frac{1}{\tfrac{4}{a^2}\sinh^2(\frac{\gamma a \tau}{2}) - 4R^2 \sin^2(\frac{\gamma \Omega \tau}{2})} - \frac{1}{ \tau^2}\right)\,.
\end{multline}

We would like to find a power series of the response function in $E$. However, the integrand is not yet sufficiently regular to do this. To begin, we may use a dominated convergence argument to take the limit $E\rightarrow 0$ under the integral to give the leading-order contribution,
\begin{align}\nonumber
    \Gamma_0 ~\coloneqq~ \lim_{E\to 0}&\F_{\tc{h}}^{\text{corr}}(E)\,,\\ ~=~ - \frac{1}{4\pi^2}\int_\RR \dd \tau\, &\left(\frac{1}{\tfrac{4}{a^2}\sinh^2(\frac{\gamma a \tau}{2}) - 4R^2 \sin^2(\frac{\gamma \Omega \tau}{2})}- \frac{1}{\tau^2}\right)\,.\label{gamma0}
\end{align}
Adding and subtracting $\Gamma_0$ to $\F_{\tc{h}}^{\text{corr}}(E)$~\eqref{eq:Fcorr} and factoring out $\tfrac{1}{\tau^2}$ yields
\begin{multline}
    \F_{\tc{h}}(E) ~=~  \Gamma_0 - \frac{1}{4\pi^2}\int_\RR \dd \tau\, \left(\frac{\cos(E \tau)-1}{\tau^2}\right)\\\times\left(\frac{\tau^2}{\tfrac{4}{a^2}\sinh^2(\frac{\gamma a \tau}{2}) - 4R^2 \sin^2(\frac{\gamma \Omega \tau}{2})}-1\right)\,.
\end{multline}
We may split this into two well-defined integrals by expanding the second pair of parentheses. The integral arising from the $-1$ term admits an elementary expression after a change of variables and integration by parts results in a simple $\text{sinc}$ integral,
\begin{align}
    \frac{1}{4\pi^2}\int_\RR\dd \tau\,\frac{\cos(E \tau)-1}{\tau^2} &~=~
    -\frac{|E|}{4\pi^2}\int_\RR\dd\sigma\,\frac{\sin(2\sigma)}{\sigma}\,\\&~=~- \frac{|E|}{4\pi}\,.
\end{align}
This result can be combined with the inertial contribution $\F_0(E)$, giving $-E/4\pi$, so that the full transition rate can now be cast as
\begin{multline}\label{fcorr small e}
    \F_{\tc{h}}^{\text{corr}}(E) ~=~  \Gamma_0 
 - \frac{E}{4\pi} \\- \frac{1}{4\pi^2}\int_\RR \dd \tau\, \frac{\cos(E \tau)-1}{\tfrac{4}{a^2}\sinh^2(\frac{\gamma a \tau}{2}) - 4R^2 \sin^2(\frac{\gamma \Omega \tau}{2})}\,.
\end{multline}
The remaining integral in Eq.~\eqref{fcorr small e} exhibits exponential decay due to the hyperbolic sine function in the denominator. Assuming $|E|<\gamma a$, we may hence Maclaurin expand the cosine and interchange the sum and the integral, obtaining
\begin{multline}
    \int_\RR \dd \tau\, \frac{\cos(E \tau)-1}{\tfrac{4}{a^2}\sinh^2(\frac{\gamma a \tau}{2}) - 4R^2 \sin^2(\frac{\gamma \Omega \tau}{2})}\\ =  \sum_{n=1}^\infty \frac{(-1)^nE^{2n}}{(2n)!} \int_\RR \dd \tau\, \frac{\tau^{2n}}{\tfrac{4}{a^2}\sinh^2(\frac{\gamma a \tau}{2}) - 4R^2 \sin^2(\frac{\gamma \Omega \tau}{2})}\,.
\end{multline}
We then define the following countable family of integrals with dimensions of $E^{1-2n}$:
\begin{equation}\label{gamman}
    \Gamma_n ~=~ -\frac{1}{4\pi^2}\int_\RR \dd \tau\, \frac{\tau^{2n}}{\tfrac{4}{a^2}\sinh^2(\frac{\gamma a \tau}{2}) - 4R^2 \sin^2(\frac{\gamma \Omega \tau}{2})}\,,
\end{equation} which gives us a series expansion for $\F_\tc{h}(E)$,
\begin{equation}\label{eqn: small E series}
    \F_\tc{h}(E) ~=~ \Gamma_0 - \frac{E}{4\pi} +  \sum_{n = 1}^\infty \frac{(-1)^n}{(2n)!} \Gamma_n E^{2n}\,,
\end{equation}convergent for $|E|<\gamma a$.
We can then directly compute an effective temperature in this limit by inverting~\eqref{detailed balance},
\begin{equation}
    T ~=~ \frac{E}{\ln\left(\frac{\F(-E)}{\F(E)}\right)}\,.
\end{equation}
Using the expansion~\eqref{eqn: small E series}, we then obtain the following small gap limit expansion for the hypertor effective temperature
\begin{equation}\label{temp small e ht}
    T_\tc{h}~=~2\pi \Gamma_0 + \left(2\pi \Gamma_1 
 - \frac{1}{24\pi\Gamma_0}\right)E^2+\mathcal{O}(|E|^4)\,.
\end{equation}

\subsection{Interpretation}

We may also write Eq.~\eqref{eqn: small E series} by means of a generating function, which is related to the case of a detector undergoing hypertor motion with Gaussian switching. The transition probability in this case reads{
\begin{align}\nonumber
    P_\tc{h}(E) &~=~ \lambda^2 T \int_\RR \dd \tau\, \ee^{- \pi \tau^2/4T^2} \ee^{-\ii E \tau} W_\tc{h}(\tau)\,,\\\nonumber &~=~ - \frac{\lambda^2 T}{4\pi^2}\int_\RR \dd\tau\, \ee^{- \pi \tau^2/2T^2}\ee^{-\ii E\tau}\\\times&\frac{1}{{\tfrac{4}{a^2}\sinh^2(\frac{\gamma a}{2}(\tau-\ii\varepsilon)) - 4R^2 \sin^2(\frac{\gamma \Omega}{2}(\tau-\ii\varepsilon))}}\,.
\end{align}}
We can rewrite $P_\tc{h}(E)$ in terms of a regular integral by adding and subtracting the vacuum transition rate multiplied by the appropriate dimensional factors, resulting in
\begin{widetext}
\begin{align}
    P_\tc{h}(E) &~=~ - \frac{\lambda^2 T}{4\pi^2}\int_\RR \dd \tau\, \cos(E \tau)\left(\frac{\ee^{- \pi \tau^2/2T^2}}{{\tfrac{4}{a^2}\sinh^2(\frac{\gamma a \tau}{2}) - 4R^2 \sin^2(\frac{\gamma \Omega \tau}{2})}}-\frac{1}{\tau^2}\right) - \lambda^2 T \int_\RR \dd \tau \,\frac{\ee^{-\ii E\tau}}{4\pi^2 (\tau-\ii\varepsilon)^2}\,,\\
    &~=~ - \frac{\lambda^2 T}{4\pi^2}\int_\RR \dd \tau\, \cos(E \tau)\left(\frac{\ee^{- \pi \tau^2/2T^2}}{{\tfrac{4}{a^2}\sinh^2(\frac{\gamma a \tau}{2}) - 4R^2 \sin^2(\frac{\gamma \Omega \tau}{2})}}-\frac{1}{\tau^2}\right) - \frac{\lambda^2 E T }{2\pi} \Theta(-E)\,,
\end{align}    
\end{widetext}where we have again defined a regular integral added to the vacuum contribution, which can be easily evaluated.

Defining $g = \pi/2T^2$, we can differentiate with respect to $g$ to obtain the coefficients $\Gamma_n$~\eqref{gamman}. Specifically, we define the time-dependent transition rate $F_\tc{h}(E;g)$ as
\begin{align}
    &F_\tc{h}(E;g) ~=~ \frac{P_\tc{h}(E)}{\lambda^2 T}\,,\nonumber\\\nonumber &\quad\quad\quad\,\,\,\,\,~=~ -\frac{1}{4\pi^2}\int_\RR \dd \tau\, \cos(E \tau)\\\times&\left(\frac{\ee^{- g\tau^2}}{{\tfrac{4}{a^2}\sinh^2(\frac{\gamma a \tau}{2}) - 4R^2 \sin^2(\frac{\gamma \Omega \tau}{2})}}-\frac{1}{\tau^2}\right) - \frac{E}{2\pi} \Theta(-E).
\end{align}
Differentiation with respect to $g$ will bring down factors of $\tau^2$ and yield regular integrals, so that for $n\geq 1$, we have
\begin{align}\nonumber
    \left.\dv{^nF_\tc{h}}{g^n}\right|_{g = 0,E=  0} &~=~\\ \frac{1}{4\pi^2 }\int_\RR& \dd \tau\, \frac{(-1)^n\tau^{2n}}{\tfrac{4}{a^2}\sinh^2(\frac{\gamma a \tau}{2}) - 4R^2 \sin^2(\frac{\gamma \Omega \tau}{2})}\,\\ &~=~ (-1)^n \Gamma_n\,.
\end{align}
Notice that evaluation at $g = 0$ corresponds to $T\to \infty$. We can, therefore, use the finite time transition rate in the limit of long interactions to write $\F_\tc{h}(E)$ in terms of derivatives of $F_\tc{h}(E;g)$ with respect to $g$ as
\begin{equation}\label{small e taylor expand}
    \F_{\tc{h}}(E) ~\sim~F_{\tc{h}}(0;0)-\frac{E}{4\pi} + \sum_{n = 1}^\infty \frac{E^{2n}}{(2n)!} F_{\tc{h}}^{(n)}(0;0)\,.
\end{equation}
The expression above is again the Taylor series given by~\eqref{eqn: small E series}, however the coefficients---which are unique whenever the Taylor series exists---are exactly given in terms of the transition probability for a detector in hypertor motion with Gaussian switching. In particular, these coefficients are given by derivatives with respect to the inverse of the interaction time, whose evaluation at $g=0$ links the behaviours of long interaction times ($g\rightarrow0^+\iff T\rightarrow\infty$)  with that of small energy gaps.

\subsection{Small-radius limit}\label{sec:small R HT}

In this subsection we consider the limit of small $R$, keeping all other parameters fixed. Intuitively, this trajectory corresponds to a small circular deviation from uniform accelerated motion. As expected, we recover the usual Unruh effect to leading order, with corrections due to the circular motion at second order. 

In studying the behaviour of $\F_\tc{h}(E)$, it will be practical to work with the following dimensionless quantities
\begin{equation}\label{dimless}
    \rho ~=~ a R\,,~\alpha ~=~ a/\Omega\,,~z ~=~ \Omega \gamma \tau/2\,, ~\varpi = \frac{2E}{\gamma \Omega}\,,
\end{equation}
so that $E\tau = \varpi z$, $a\gamma \tau/2 = \alpha z$, $R\Omega = \rho/\alpha$, and the small-radius limit is the limit of small $\rho$. In these variables, we then define
\begin{align}\nonumber
    w_\tc{h}(z) &~\coloneqq~ W_{\tc{h}}(2z/\gamma \Omega )\,, \\&~=~ \frac{\alpha^2 \Omega^2}{16\pi^2(\rho^2 \sin^2(z-\ii\varepsilon) - \sinh^2(\alpha (z-\ii\varepsilon)))}\,.
\end{align}
\noindent This leads to the following transition rate,
\begin{align}
    \F_{\tc{h}}(E) &~=~ \frac{2}{\gamma \Omega}\int_\RR \dd z\, \ee^{-\ii\varpi z} w_{\tc{h}}(z)\,\\\nonumber &~=~ \frac{\alpha^2 \Omega}{8\pi^2 \gamma }\int_\RR \dd z\, \ee^{-\ii\varpi z}\\&\frac{1}{\rho^2 \sin^2(z - \ii \varepsilon) - \sinh^2(\alpha( z - \ii \varepsilon))}\,.\label{dimless F}
\end{align}
\noindent As before, we isolate the distributional behaviour of $\F_\tc{h}(E)$ and split it into the \textit{inertial} and \textit{correction} contributions. The inertial term is again given by Eq.~\eqref{minkowski response} with the correction taking the following form,
\begin{multline}\label{integrand complex}
     \F_{\tc{h}}^\text{corr}(E) ~=~ - \frac{\gamma \Omega}{4 \pi^2}\int_0^\infty \dd z\, \cos(\varpi z)\\\times \left(\frac{\alpha^2 - \rho^2}{\sinh^2(\alpha z) - \rho^2 \sin^2(z)} - \frac{1}{z^2}\right)\,.
\end{multline}
The timelike condition $R\Omega<1$ corresponds in these variables to $\rho<\alpha$. A dominated convergence argument allows us to take the limit $\rho\rightarrow0$ under the integral, leaving the leading-order contribution
\begin{multline}
    \lim_{\rho\rightarrow0}\F_{\tc{H}}^{\mathrm{corr}}(E)~=~-\frac{\gamma\Omega}{4\pi^2}\int_0^\infty\dd z\,\cos(\varpi z)\\\times\left(\frac{\alpha^2}{\sinh^2(\alpha z)}-\frac{1}{z^2}\right)\,.
\end{multline}
We look now at the subleading contributions, which can be conveniently rewritten as
\begin{multline}
    \F_{\tc{H}}^{\mathrm{corr}}(E)-\lim_{\rho\rightarrow0}\F_{\tc{H}}^{\mathrm{corr}}(E)~=~-\rho^2\frac{\gamma\Omega}{4\pi^2}\\\times\int_0^\infty\dd z\,\underbrace{\cos(\varpi z)\frac{\left(\frac{\alpha^2\sin^2(z)}{\sinh^2(\alpha z)}-1\right)}{\sinh^2(\alpha z)}}_{\textbf{(a)}}\underbrace{\frac{1}{1-\rho^2\frac{\sin^2(z)}{\sinh^2(\alpha z)}}}_{\textbf{(b)}}\,.
\end{multline}Function \textbf{(a)} is independent of $\rho$ and is a regular, integrable function of $z$. Function \textbf{(b)} can be expanded to the first $n$ terms as a geometric series with a remainder term. The remainder term is given by
\begin{multline}
    \rho^2\int_0^\infty\dd z\,\cos(\varpi z)\frac{\left(\frac{\alpha^2\sin^2(z)}{\sinh^2(\alpha z)}-1\right)}{\sinh^2(\alpha z)}\left(\rho^2\frac{\sin^2(z)}{\sinh^2(\alpha z)}\right)^{n+1}\\\times\frac{1}{1-\rho^2\frac{\sin^2(z)}{\sinh^2(\alpha z)}}\,.
\end{multline} We can bound $(1-\rho^2\sin^2(z)/\sinh^2(\alpha z))^{-1}$ above by $(1-\rho^2/\alpha^2)^{-1}$. The remaining integral in this bound then converges due to the exponential suppression at infinity and regularity at zero. Hence, the remainder term is order $\rho^{2n+4}$. We are interested in the small-$\rho$ limit so we can assume $\rho<1$ and the remainder, therefore, tends to zero as $n\rightarrow\infty$. This justifies the following expansion in small $\rho$ under the integral,
\begin{multline}\label{hypertor small R expand}
    \F_{\tc{h}}^{\text{corr}}(E) ~=~ -\frac{\gamma \Omega}{4 \pi^2}\int_0^\infty \dd z\, \cos(\varpi  z) \left(\frac{\alpha^2}{\sinh^2(\alpha z)} - \frac{1}{z^2}\right)\\+\frac{\gamma \Omega\rho^2}{4 \pi^2}\int_0^\infty \dd z\, \cos(\varpi  z) \left(\frac{1}{\sinh^2(\alpha z)} - \frac{\alpha^2 \sin^2(z)}{\sinh^4(\alpha z)}\right)\\ + \mathcal{O}(\rho^4).
\end{multline}
Both integrals in Eq.~\eqref{hypertor small R expand} admit elementary closed forms. We perform these integrals and revert to dimensionful quantities, which yields the leading-order corrections in $\F(E)$ for small radii,
\begin{widetext}
\begin{multline}
    \F_{\tc{h}}(E)~=~ \frac{E}{2\pi}\frac{1}{e^{2\pi E/a} - 1}+\frac{aR^2}{24 \pi\gamma}\left(\left[\gamma^2 a^2+ (E-\gamma\Omega)^2\right]g(\tfrac{2(E-\gamma\Omega)}{\gamma a}) \right. - 2\left[3\gamma^2\Omega^2 + \gamma a^2 + E^2\right]g(\tfrac{E}{\gamma a})\\+\,\,\,\left.\left[\gamma^2 a^2 +(E+\gamma\Omega)^2\right]g(\tfrac{2(E + \gamma \Omega)}{\gamma a})\right) + \mathcal{O}(R^4)\,,\label{eqn: F ht small R}
\end{multline}    
\end{widetext}where $g(u) = u \coth(\pi u)$. To leading order, we recover the characteristic thermal response of the Unruh effect. {The subleading term is more complicated; however, since the function $g(u)$ is even, this subleading term is then an even function of $E$.}

\subsection{Large-gap limit}\label{sec: large gap ht}
We consider in this section the large detector gap limit (large $\varpi$), with all other parameters fixed. Given the usual split of the response function into its inertial and correction contributions, we need only focus on the latter. We will analyze this using the form in Eq.~\eqref{integrand complex},
\begin{multline}\label{eqn: Fcorr large E2}
    \mathcal{F}_{\tc{h}}^{\text{corr}}(E)~=~\frac{\Omega\gamma}{8\pi^2}\int_\RR\dd z \,\cos(\varpi z)\\\times\left(\frac{1}{z^2}+\frac{\alpha^2-\rho^2}{\rho^2\sin^2(z)-\sinh^2(\alpha z)}\right)\,.
\end{multline}In~\eqref{eqn: Fcorr large E2}, we first replace $\cos(\varpi z)$ by $\exp(\ii|\varpi|z)$, by the evenness of the integrand. We then deform the integration contour from the real axis to a contour $\mathcal{C}$ that passes the pole at $z=0$ in the upper half-plane, on (say) a semicircle so small that the contour deformation crosses no singularities. The residue theorem informs us that the integral $\int_{\mathcal{C}}\exp(\ii|\varpi|z)z^{-2}\,\dd z$ vanishes, leaving us with

\begin{equation}\label{eqn: Fcorr large E}
    \mathcal{F}_{\tc{h}}^{\text{corr}}(E)~=~\frac{\Omega\gamma}{8\pi^2}\int_{\mathcal{C}}\dd z \,\ee^{\ii|\varpi|  z}\frac{\alpha^2-\rho^2}{\rho^2\sin^2(z)-\sinh^2(\alpha z)}\,.
\end{equation}A contour integration argument then says that the leading contribution to~\eqref{eqn: Fcorr large E} at large $|\varpi|$ comes from the pole in the upper half-plane that is closest to the real axis. We need to identify this pole.

We consider the first purely imaginary pole of the integrand, $z=\ii\mu$ with $\mu>0$. Such a pole would satisfy
\begin{equation}
    \sin^2(\alpha\mu)-\rho^2\sinh^2(\mu)~=~0\,.
\end{equation} Given $\mu,~\alpha,~\rho>0$, this simplifies to
\begin{equation}\label{eqn: mu def}
    \sin(\alpha\mu)~=~\rho\sinh(\mu)\,.
\end{equation} Due to the ordering $\alpha>\rho$, this has at least one nonzero solution, the smallest of which we take to define $\mu$. One may then question whether there are any poles off of the imaginary axis with positive imaginary part less than or equal to $\mu$. In order to address this question, it is possible to check by considering $z=x+\ii Y$ as a function of $x$ with $x\in\mathbb{R}\setminus\{0\}$ and $0<Y\leq\mu$ fixed. {We have confirmed that the smallest pole with positive imaginary part is indeed given by $z=\ii\mu$.}

An application of the residue theorem then gives the leading contribution to $\mathcal{F}^{\text{corr}}(E)$ in the large energy gap limit as
\begin{equation}\label{largegaptemphyper}
    \mathcal{F}_{\tc{h}}^{\text{corr}}(E)~\sim~\frac{\alpha a}{4\pi\gamma}\frac{\ee^{- \mu | \varpi |}}{\rho^2\sinh(2\mu)-\alpha\sin(2\alpha\mu)}\,, \text{ as } |\varpi|\rightarrow\infty.
\end{equation}

One may consider then the effective temperature derived from this response. As is typical with large-gap expansions~\cite{Biermann:2020bjh}, the leading contribution to the effective temperature is given by the coefficient of the exponent that multiplies $|E|$. Recalling that $\mu |\varpi| = \frac{2 \mu}{\Omega \gamma} |E|$, we obtain
\begin{equation}
    T_\tc{h}~=~\frac{\gamma\Omega}{2\mu}+o(1)\,,\text{ as }|E|\rightarrow\infty.
\end{equation}

\section{Circular Motion in a Thermal Bath}\label{sec:centerOfMass}

In this section, we summarize our results for a similar analysis of a UDW detector in a $3+1$ thermal bath undergoing circular motion. We assume the circular motion to have no drift in the rest frame of the thermal bath. In this case, the system is time-translation invariant along the trajectory~\cite{bunnyCircular}. We analyze the same limits as for the hypertor motion, except for the small-radius limit, whose analysis we defer until Section~\ref{sec:versus}.

The circular motion trajectory in Minkowski spacetime can be parametrized by proper time as
\begin{equation}\label{circ traject}
    \mf z (\tau) ~=~ (\gamma \tau,  R \cos(\Omega \gamma \tau),  R \sin(\Omega \gamma \tau),0)\,,
\end{equation} which is the integral curve of the combination of a rotational Killing vector field and a time-translation vector field of Minkowski spacetime. In Eq.~\eqref{circ traject}, $R$ represents the radius trajectory as witnessed by an observer undergoing inertial motion in the center of the circular trajectory, and $\Omega$ is the angular velocity measured by the same observer. The gamma factor is then again given by~\eqref{eqn: gamma def}.

We consider a particle detector undergoing this motion in Minkowski spacetime while interacting with a quantum scalar field initially prepared in a thermal state at temperature $T$. We define the parameter $a = 2\pi T$, so that the pullback of the Wightman function and the associated response function are written as
\begin{widetext}
    \begin{align}\label{HT Wightman}
    W_{\tc{tb}}(\tau) &~=~ -\frac{1}{4\pi^2}\frac{a \sinh(2 a R \sin(\tfrac{\gamma\Omega}{2}\tau))}{4 R \sin(\tfrac{\gamma \Omega}{2}\tau)\left\{\cosh(a\gamma (\tau-\ii\varepsilon)) - \cosh(2a R \sin(\tfrac{\gamma \Omega}{2}(\tau-\ii\varepsilon)))\right\}}\,,\\
    \F_{\tc{tb}}(E) &~=~ \int_\RR \dd \tau\, \ee^{- \ii E\tau}W_{\tc{tb}}(\tau)\,.\label{TB response}
\end{align}
\end{widetext}We define the value of $f(\tau) := \sinh(2aR\sin(\gamma\Omega\tau/2))/\sin(\gamma\Omega\tau/2)$ at all zeros of $\sin(\gamma\Omega\tau/2)$ to be the value of the limit $\lim_{\tau\rightarrow0}f(\tau)=2aR$, rendering the Wightman function~\eqref{HT Wightman} well-defined. The parameter $a$ is in principle a simple rescaling of the temperature $T$, but defined such that a particle undergoing uniform linear acceleration $a$ would experience a thermal bath at the temperature $T$. In particular, we have{
\begin{equation}
    \lim_{R\to 0} W_\tc{tb}(\tau) ~=~  - \frac{1}{4\pi^2}\frac{1}{\frac{4}{a^2}\sinh^2(\tfrac{a}{2}(\tau-\ii\varepsilon))}\,.
\end{equation}}
This is the correlation function seen by a uniformly accelerated observer with proper acceleration $a$ (see Eq.~\eqref{accel Wightman}), or equivalently, by a static observer in a thermal bath (Eq.~\eqref{thermal bath wightman}).

As in the case of hypertor motion discussed in Section~\ref{sec:hypertor}, the integral defining the detector's response cannot be evaluated in terms of elementary functions, and we thus explore Eq.~\eqref{TB response} in limiting regimes. The techniques used in this section are very similar to the ones used in Section~\ref{sec:hypertor}, or have been studied in past literature~\cite{bhDetectors,Biermann:2020bjh}, so we omit most of the calculation details in this section. We also remark that we postpone discussing the limit of a small orbital radius to Section~\ref{sec:versus}.

\subsection{Small-gap limit}\label{subsec: small tb}
The short-distance behaviour of the Wightman function in a thermal bath is also of Hadamard form,{
\begin{equation}\label{hadamard}
    W_{\tc{tb}}(\tau) ~\sim~ -\frac{1}{4\pi^2 (\tau-\ii\varepsilon)^2}\,,\text{ as }\tau\rightarrow0\,.
\end{equation} }
{We again utilize the techniques of~\cite{handsome}, as in Subsection~\ref{subsec: small gap HT}, to isolate the distributional contribution arising from the short-distance behaviour~\eqref{hadamard} and obtain a small-$E$ expansion of the response function akin to~\eqref{eqn: small E series}. 
Furthermore, the term $\cosh(a\gamma(\tau-\ii\varepsilon))$ forces exponential convergence of the integral coefficients in the expansion. As such, we obtain}{
\begin{equation}\label{small e tb}
    \F_{\tc{tb}}(E)~\sim~\tilde{\Gamma}_0-\frac{E}{4\pi}+\sum_{n=1}^\infty\frac{(-1)^n}{(2n)!}\tilde{\Gamma}_nE^{2n}\,,
\end{equation}} where $\Tilde{\Gamma}_0$ and $\Tilde{\Gamma}_n$ are analogous to hypertor small energy gap expansion coefficients $\Gamma_0$~\eqref{gamma0} and $\Gamma_n$~\eqref{gamman},
\begin{widetext}
    \begin{align}\label{gamma0 tilde}
    \tilde{\Gamma}_0~=~& - \frac{1}{4\pi^2}\int_\RR \dd \tau\, \left(\frac{a \sinh(2 a R \sin(\Omega \gamma \tau/2))}{4 R \sin(\gamma \Omega \tau/2)\{\cosh(a\gamma \tau) - \cosh(2a R \sin(\gamma \Omega \tau/2))\}} - \frac{1}{\tau^2}\right)\,,\\
    \tilde{\Gamma}_n~=~&- \frac{1}{4\pi^2}\int_\RR \dd \tau\, \tau^{2n}\left(\frac{a \sinh(2 a R \sin(\Omega \gamma \tau/2))}{4 R \sin(\gamma \Omega \tau/2)\{\cosh(a\gamma \tau) - \cosh(2a R \sin(\gamma \Omega \tau/2))\}}\right)\,.\label{gamman tilde}
\end{align}
\end{widetext}One may note the same order $E$ term present in both Eq.~\eqref{eqn: small E series} and Eq.~\eqref{small e tb}. The remaining contributions are given by the $\tilde{\Gamma}_n$ terms, wherein any deviations between $\F_\tc{tb}(E)$ and $\F_\tc{h}(E)$ may arise. 

We can immediately write down the small-gap expansion for the effective temperature for circular motion in a thermal bath,
\begin{equation}\label{temp small e tb}
    T_{\tc{tb}}~=~2\pi\tilde{\Gamma}_0+\left(2\pi \Tilde{\Gamma}_1 -\frac{1}{24 \pi \tilde{\Gamma}_0}\right)E^2+\mathcal{O}(|E|^4)\,.
\end{equation}

\subsection{Large-gap limit}\label{subsec: large e tb}
The limit of large energy gap for circular motion in a thermal bath has already been analyzed in~\cite{bhDetectors}; hence, we report the relevant findings briefly for later comparison. It is again practical to work with the dimensionless quantities of Eq.~\eqref{dimless} as per Subsections~\ref{sec:small R HT} and~\ref{sec: large gap ht}, and we define
\begin{align}\nonumber
    w_{\tc{tb}}(z) &~=~ W_{\tc{tb}}(2z/\gamma \Omega)\,,\\ &~=~ \frac{\alpha^2 \Omega^2 \csc(z)\sinh(2 \rho \sin z)}{16\pi^2 \rho(\cosh(2\rho \sin z) - \cosh(2 \alpha z))}\,,
\end{align}
such that the correction to the inertial motion transition rate is given by
\begin{multline}\label{tb correction}
    \F^{\text{corr}}_{\tc{tb}}(E) ~=~ \frac{\gamma \Omega}{8 \pi^2}\int_\RR\dd z\, \cos(\varpi z) \\\times\left(\frac{1}{z^2} + \frac{(\alpha^2 - \rho^2)\csc(z)\sinh(2\rho \sin z)}{\rho 
 (\cosh(2\rho \sin z) - \cosh(2 \alpha z))}\right)\,.
\end{multline}

We consider now the limit $|E|\rightarrow\infty$ ($\varpi \to \infty$) with all other parameters fixed. Performing the extension to the complex plane detailed in~\cite{jorma}, one finds the leading contribution to the correction response in~\eqref{tb correction} is given by the nonzero pole with the smallest positive imaginary part. However, the singularity structure is more complicated here than in the case of hypertor motion due to the presence of a critical temperature value $T_{\text{crit}}$, characterizing a change in the position of this pole. Defining the critical parameter $a_{\text{crit}} = 2\pi T_{\text{crit}}$, $a_{\text{crit}}$ is obtained as a solution to the transcendental equation
\begin{equation}
    R \Omega = \frac{2R a_{\text{crit}}}{\pi} \text{arcsinh}\left(\frac{\pi}{2Ra_{\text{crit}}}\right).
\end{equation}
The pole with the smallest positive imaginary part is given by $z=\ii\mu_-$ for $a>a_{\text{crit}}$ and by $z=\ii\mu_+$ for $a<a_{\text{crit}}$. It was shown~\cite{bhDetectors} that the poles are the solutions to the following transcendental equations
\begin{equation}
    \frac{\mu_+}{R\Omega}~=~\sinh\mu_+\,, \quad \quad \frac{1}{2 R T} ~=~  \frac{\mu_-}{R\Omega} + \sinh\mu_-\,.
\end{equation}
In the limit of large energy gaps, we have
\begin{multline}\label{eqn: large E rate circ}
    \F_{\tc{tb}}^{\text{corr}}(E)~\sim~\frac{1}{8\pi R\gamma}
    \frac{\ee^{-|\varpi|\mu_\pm}}{\sinh \mu_\pm(R\Omega\cosh\mu_\pm\mp1)}\,,\\\text{ as }|E|\rightarrow\infty\,,
\end{multline}
\noindent where the choice of $\mu_\pm$ is determined by whether $a>a_{\text{crit}}$ or $a<a_{\text{crit}}$.
The resulting effective temperature experienced by the detector is then given by the coefficient in the exponential, $|\varpi| \mu_\pm = \frac{2 \mu_\pm}{\gamma \Omega}|E|$. That is,
\begin{equation}\label{templargegap}
    T_{\tc{tb}}~=~\frac{\gamma\Omega}{2\mu_\pm}+o(1)\,,\text{ as }|E|\rightarrow\infty\,.
\end{equation}

\section{Thermal Bath vs. Uniform Acceleration}\label{sec:versus}

In this section, we compare the case of a UDW detector undergoing hypertor motion in the vacuum with the case of circular motion in a thermal bath. We contrast the two motions in the limiting regimes explored in Sections~\ref{sec:hypertor} and~\ref{sec:centerOfMass} and provide numerical plots to cover a wide range of the parameter space.

\subsection{Small-gap limit}
We begin our comparison with the case of small energy gaps, as studied in Subsections~\ref{subsec: small gap HT} and~\ref{subsec: small tb}. Given that the integral expressions for the response functions in the two cases were amenable to the same techniques, one notes that the expansions in Eq. ~\eqref{eqn: small E series} and Eq.~\eqref{small e tb} have the same form, albeit with different coefficients. The small energy gap behaviour of the transition rate is determined by the behaviour of the integrals $\Gamma_0$~\eqref{gamma0} and $\tilde{\Gamma}_0$~\eqref{gamma0 tilde}. These coefficients also determine the leading-order behaviour of the temperatures of the detectors (Eqs. ~\eqref{temp small e ht} and~\eqref{temp small e tb}) in this regime. As such, we now direct our attention towards comparing $\Gamma_0$ and $\tilde{\Gamma}_0$. 

\begin{figure*}
\begin{subfigure}{.45\textwidth}
  \centering
  \includegraphics[width= 8cm]{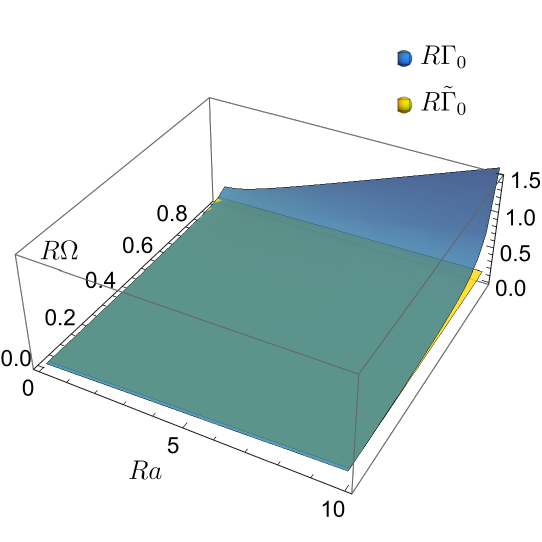}
  \caption{}
    \label{fig:Gamma0legends}
\end{subfigure}
\begin{subfigure}{.45\textwidth}
  \centering
  \includegraphics[width=8cm]{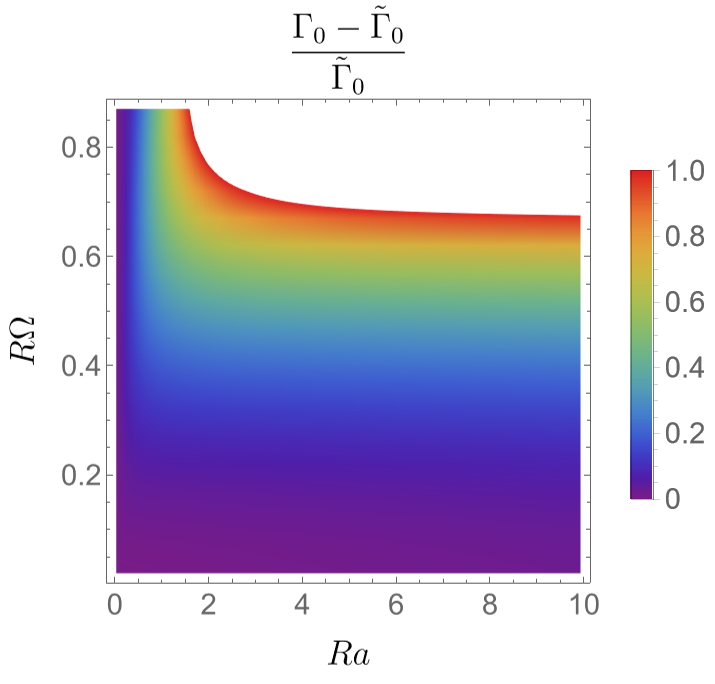}
    \caption{}
    \label{fig:Gamma0relative}
\end{subfigure}%
\caption{\justifying Comparison of the leading-order coefficients in the expansion of $\F(E)$ for detectors undergoing hypertor motion ($\Gamma_0$) and circular motion in a thermal bath of temperature $a/2\pi$ ($\tilde{\Gamma}_0$), where $a$ is the uniform acceleration parameter of the hypertor. \textbf{(a)} $R\Gamma_0$ and $R\tilde{\Gamma}_0$ as a function of $Ra$ and $R\Omega$ with plotting range $0\leq R\Omega\leq 0.91$ and $0\leq Ra\leq 10$. \textbf{(b)} The normalized relative difference between $\Gamma_0$ and $\tilde{\Gamma}_0$ as a function of $Ra$ and $R\Omega$ with plotting range $0\leq R\Omega\leq 0.86$ and $0\leq Ra\leq 10$. The white region represents when the normalized relative difference is greater than one.}
\end{figure*}

We investigate these differences numerically. In Fig.~\ref{fig:Gamma0legends}, we use the orbital radius as a characteristic length scale and plot $R\Gamma_0$ and $R\tilde{\Gamma}_0$ as functions of the dimensionless variables $Ra$ and $R\Omega$. For trajectories with low circular motion speeds, the two coefficients match. This is expected as this limit approaches the case of either a uniformly linearly accelerated detector, or a static detector in a thermal bath, the known duality due to the Unruh effect. Furthermore, the two coefficients also match in the limit of small accelerations/low initial thermal state temperatures. This is also expected as both scenarios reduce to uniform circular motion interacting with the Minkowski vacuum in $3+1$ Minkowski spacetime. We find that the leading-order contribution to $\F_\tc{h}(E)$ ($T_\tc{h}$) is always larger than the leading-order contribution to $\F_\tc{tb}(E)$ ($T_\tc{tb}$). The discrepancy between the two cases grows as $R\Omega\rightarrow 1$, in which cases both coefficients diverge.

Overall, one sees that particle detectors with small energy gaps would be able to distinguish between circular motion interacting with a field initially prepared in a thermal state at temperature $a/2\pi$ and a hypertor trajectory interacting with the Minkowski vacuum. The detectors are enabled to distinguish the two cases more clearly in the case where both $a$ and the angular velocity $\Omega$ are large compared to $1/R$.
In Fig.~\ref{fig:Gamma0relative}, we plot the normalized relative difference between $\Gamma_0$ and $\tilde{\Gamma}_0$. This highlights the regions of parameter space where the two coefficients, and hence two motions, are most dissimilar. For small accelerations/low initial state temperatures, there is a region where the two coefficients differ by less than $10\%$. Furthermore, for low speeds $R\Omega\lesssim0.1$, one may notice how similar the two motions are.

\subsection{Small-radius limit}

In this subsection, we compare the experience of UDW detectors undergoing hypertor motion and circular motion in a thermal bath in the limit of small orbital radii. As established, the short-distance behaviours of the two trajectories are identical and hence the difference in their Wightman functions is itself a regular function. We, therefore, utilize this in comparing the transition rates of the two motions. We write the difference in transition rates as
\begin{align}\label{eqn: delta F small R}\nonumber
    \Delta \F(E) &~=~ \F_{\tc{h}}(E) - \F_\tc{tb}(E)\,, \\&~=~ \int_\RR \dd \tau\, \ee^{- \ii E \tau} (W_{\tc{h}}(\tau) - W_{\tc{tb}}(\tau))\,.
\end{align}
Should we be able to compute~\eqref{eqn: delta F small R}, we can then write down a closed form $\F_{\tc{tb}}(E) = \F_\tc{h}(E) - \Delta \F(E)$ in the limit of small $R$, using the results of~\eqref{eqn: F ht small R}.

For simplicity, we again use the dimensionless variables~\eqref{dimless} and write the difference in the transition as
\begin{widetext}
    \begin{equation}
    \Delta\F(E)  ~=~ \frac{\gamma \Omega}{8 \pi^2}\int_0^\infty \dd z\, \cos(\varpi z) \left(\frac{\alpha^2-\rho^2}{\rho^2\sin^2(z)-\sinh^2(\alpha z)} - \frac{(\alpha^2 - \rho^2)\csc(z)\sinh(2\rho \sin z)}{\rho 
 (\cosh(2\rho \sin z) - \cosh(2 \alpha z))}\right)\,.
\end{equation}
\end{widetext}Each of the integrals above has a single singularity at $z = 0$ over the real line, given by~\eqref{hadamard ht} and~\eqref{hadamard}. As such, these cancel each other out, leaving a regular function at $z = 0$. Moreover, the exponential decay provided by $\sinh^2(\alpha z)$ and $\cosh(2\alpha z)$ in each term respectively is enough to bound the remainder term from Taylor's theorem as in Section~\eqref{sec:small R HT} so that we can perform a small-$\rho$ expansion inside the integral. Indeed, we find that
\begin{multline}
    \frac{\alpha^2-\rho^2}{\rho^2\sin^2(z)-\sinh^2(\alpha z)} - \frac{(\alpha^2 - \rho^2)\csc(z)\sinh(2\rho \sin z)}{\rho 
 (\cosh(2\rho \sin z) - \cosh(2 \alpha z))} \\~=~ \frac{2\alpha^2 \rho^2}{3}\frac{\sin^2(\alpha z)}{\sinh^2(z)} + \mathcal{O}(\rho^4)\,.
\end{multline}
Plugging this result into Eq.~\eqref{eqn: delta F small R}, the leading-order behaviour of $\Delta\F(E)$ is given by
\begin{equation}\label{delta F1}
    \Delta\F(E) = \frac{\gamma \Omega \alpha^2 \rho^2}{12\pi}\int_0^\infty\cos(\varpi z) \frac{\sin^2(z)}{\sinh^2(\alpha z)} + \mathcal{O}(\rho^4)\,,
\end{equation} which has an elementary form, which is given to leading order in $R$ in dimensionful variables as
\begin{multline}\label{delta F2}
    \Delta \F(E) ~=~ \frac{a^3 R^2}{48 \pi}\left(g(\tfrac{E + \Omega}{a})+g(\tfrac{E - \Omega}{a}) - 2g(\tfrac{E}{a})\right)\\+\mathcal{O}(R^4)\,.
\end{multline}
where $g(u) = u \coth(\pi u)$. One may show that $g(u+v) + g(u-v) - 2 g(u)>0$, and one sees that in this regime the transition rate for the hypertor motion is larger than that for the circular motion in a thermal bath. Furthermore, since the difference is $\mathcal{O}(R^2)$, the leading-order behaviour of each motion in the small-$R$ limit is identical, which recovers the Unruh effect.

\begin{figure*}[t]
\begin{subfigure}{.49\textwidth}
  \centering
  \includegraphics[width=8.6cm]{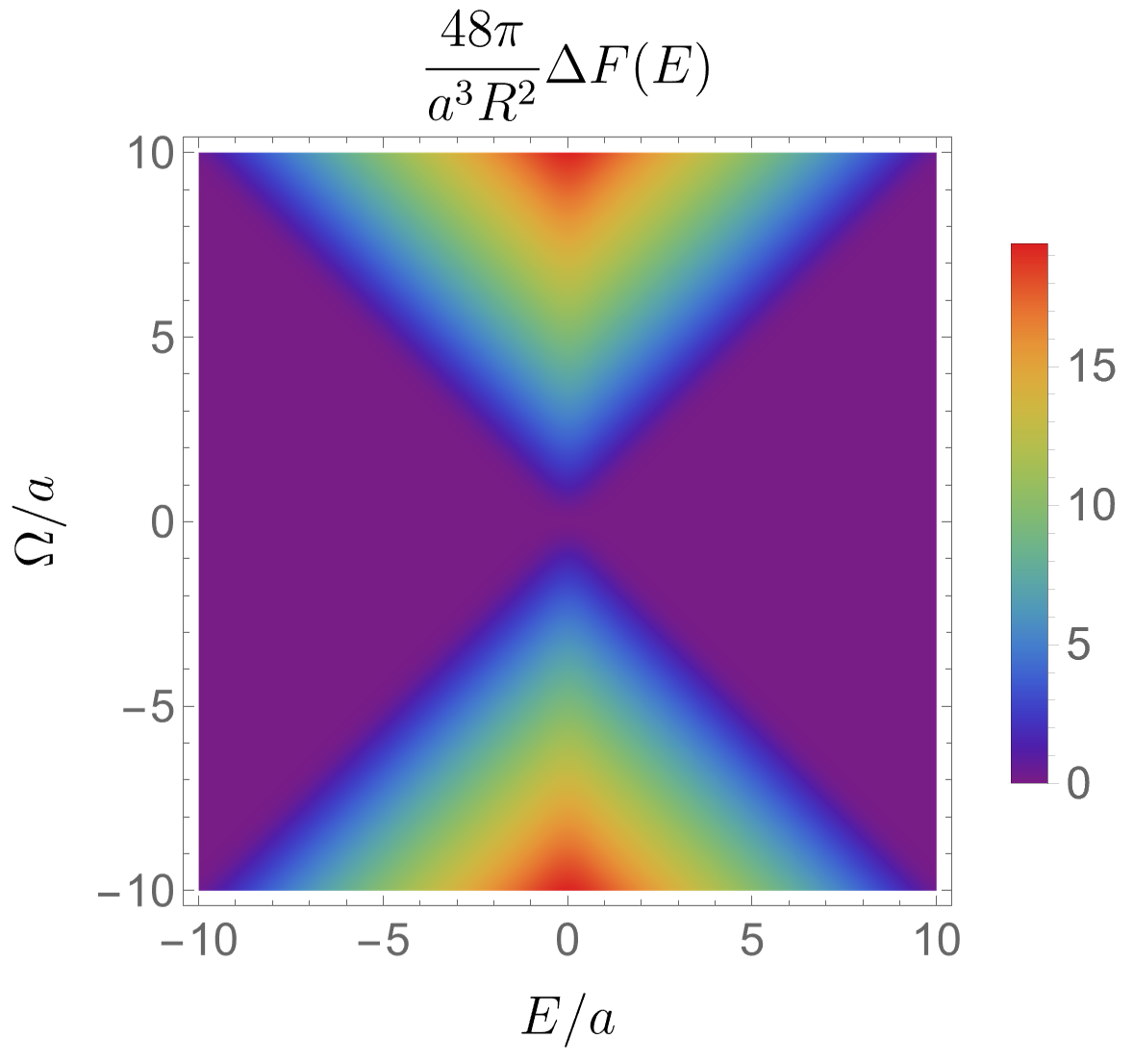}
    \caption{}
    \label{fig:PlotDeltaF}
\end{subfigure}%
\begin{subfigure}{.49\textwidth}
  \centering
  \includegraphics[width=8.6cm]{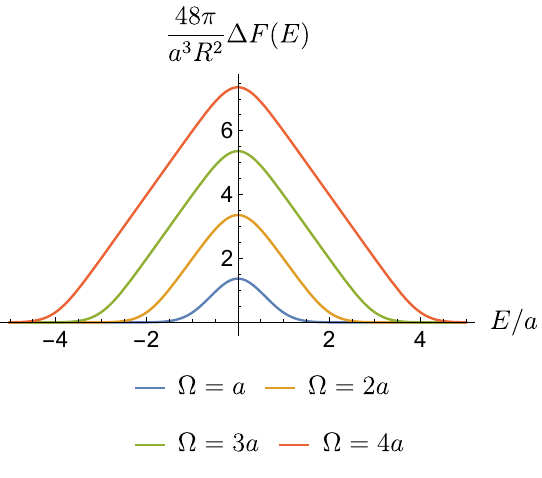}
    \caption{}
    \label{fig:PlotDeltaF1D}
\end{subfigure}
\caption{\justifying leading-order correction coefficient to the transition rate difference in Eq.~\eqref{eqn: delta F small R}. \textbf{a)} Density plot of the leading-order coefficient with plotting ranges $-10\leq E/a\leq 10$, $-10\leq \Omega/a\leq 10$. \textbf{b)} Horizontal slices from a) for fixed angular velocities with plotting range $-5\leq E/a\leq5$.}
\label{fig:DeltaF}
\end{figure*}

In Fig.~\ref{fig:DeltaF}, we plot the adimensionalized leading-order coefficient of the difference~\eqref{eqn: delta F small R}. From Fig.~\ref{fig:PlotDeltaF}, one sees clearly that for $|E|>|\Omega|$, the leading-order behaviour of the detectors is quantitatively similar. On the other hand, for $|E|\lesssim|\Omega|$, there is an appreciable difference between the two motions. This is characteristically similar to the comparison of inertial and circular motions in a $2+1$ thermal bath in~\cite{bunnyCircular}, where a detector probing above the frequency of the circular motion would not be able to distinguish between inertial and circular motion. In Fig.~\ref{fig:PlotDeltaF1D}, we plot the behaviour of the leading-order term for a fixed angular velocity $\Omega$. This highlights the property that probing below the frequency of the circular motion enables a detector to distinguish better between the two motions.

The difference in transition rates~\eqref{eqn: delta F small R} also enables us to compute the leading-order difference in effective temperatures. Given the results of Subsection~\ref{sec:small R HT}, both response functions are of the form $\F(E)=\F_a(E)+R^2\delta\F(E)$, with $\F_a(E)$ given by Eq.~\eqref{eqn:lin accel}. Furthermore, $\Delta\F(-E)=\Delta\F(E)$ due to the evenness of $g(u)$, together with the result of $\F_\tc{h}(E)$ from Subsection~\ref{sec:small R HT}. Therefore, $\delta\F(-E) = \delta\F(E)$. 

Considering a response function of the form $\F(E)=\F_a(E)+R^2\delta\F(E)$, the effective temperature can then be expanded as a series in $R$,
\begin{align}
    T&~=~\frac{E}{\ln\left(\frac{\F_a(-E)+R^2\delta\F(E)}{\F_a(E)+R^2\delta\F(E)}\right)}\,,\\&~=~T_U\nonumber\\&+R^2\delta\F(E)\left(\frac{1}{\F_a(E)}-\frac{1}{\F_a(-E)}\right)\frac{T_U}{\ln\left(\frac{\F_a(-E)}{\F_a(E)}\right)}\nonumber\\&\quad\quad\quad\quad\quad\quad\quad\quad\quad\quad\quad\quad\quad\quad\quad\quad+\mathcal{O}(R^3)\,,
\end{align}where $T_U=a/2\pi$ is the Unruh temperature and $\F_a(E)$ is the corresponding transition rate (Eq.~\eqref{detailed balance}). One may note that the correction is positive, since $\F_a(-E)\geq\F_a(E)$ for $E>0$.

We can also perform this expansion for the difference in temperatures,  $\Delta T~=~T_{\tc{h}}-T_{\tc{tb}}$. The small-radius expansion of the temperature difference can then be written as
\begin{multline}
    \Delta T
    ~=~\Delta\F(E)\left(\frac{1}{\F_a(E)}-\frac{1}{\F_a(-E)}\right)\frac{T_U}{\ln\left(\frac{\F_a(-E)}{\F_a(E)}\right)}\\+\mathcal{O}(R^3)\,.
\end{multline}
We remark that this expression does not depend on the sign of $E$ since $\Delta\mathcal{F}(E)$, given by~\eqref{delta F1}, is even and the combination $(1/\mathcal{F}_a(E)-1/\mathcal{F}_a(-E))\left(\ln(\tfrac{\mathcal{F}_a(-E)}{\mathcal{F}_a(E)})\right)^{-1}$ is even.

Thus, the temperature difference experienced by the detectors in the small-radius limit is proportional to both the Unruh temperature $T_U = a/2\pi$ and to the difference in the transition rate of the detectors. Notice that because $\Delta\F(E)$ is always positive, the hypertor transition rate is always larger than the circular motion one in the small-radius limit, then the temperature experienced by a detector undergoing hypertor motion will always be larger than in circular motion for small radius.

\vspace{-0mm}

\subsection{Large-gap limit}
We now turn our attention to the large energy gap regime to compare the two motions. {In this section, we first perform the large-gap expansions as studied in Subsections~\ref{sec: large gap ht} and~\ref{subsec: large e tb} and then ask how the leading-order term in this expansion looks in two further regimes. We show that the resulting detector responses are indistinguishable to leading order in the limits of small $Ra$ and small $R\Omega$.}

Both the circular motion in a thermal bath and hypertor motion large-gap behaviours depend on the roots of transcendental equations. We present these in full for completeness. For the hypertor motion, the pole with smallest positive imaginary part is given by $z=\ii\mu$ where $\mu$ is the first nonzero solution to
\begin{equation}
    \sin(\tfrac{Ra}{v}\mu)~=~Ra\sinh(\mu)\,,\label{transeqm}
\end{equation}where $v=R\Omega$. In the case of a detector in circular motion in a thermal bath at temperature $T$, the leading behaviour depends on whether $T>T_{\text{crit}}$ or $T<T_{\text{crit}}$. This can be summarized as
\begin{align}
       \mu_+&~=~v\sinh(\mu_+)\,,&T<T_{\text{crit}}\,,\label{transeqmp}\\
       v&~=~2RT\mu_-+2RTv\sinh(\mu_-)\,,&T>T_{\text{crit}}\,,\label{transeqmm}\\
    \frac{1}{4RT_{\text{crit}}}&~=~\sinh\left(\frac{v}{4RT_{\text{crit}}}\right)\,.\label{transeqT}
\end{align} We consider the limits of small acceleration/low initial thermal state temperature and small velocity. In this case, one expects the two motions to tend to $3+1$ circular motion in the Minkowski vacuum. In the case of small velocities, one expects the hypertor motion and circular motion through a $3+1$ thermal bath to tend to an accelerated trajectory and a static detector in a $3+1$ thermal bath respectively, being equivalent due to the Unruh effect.

\vspace{-0mm}

\subsubsection{Limit of small acceleration parameter}\label{subsubsec accel}
We consider first the limit of small parameter $a$, representing either small perpendicular acceleration, or a low initial temperature for the thermal state. If we consider the defining equation for $\mu$ \eqref{transeqm} in the pointwise limit of $a \rightarrow0$ and look for a nonzero, positive solution, we find that~\eqref{transeqm} is equivalent to
\begin{equation}\label{pointwise limit}
    \mu~=~v\sinh(\mu)\,.
\end{equation} This has three solutions for $0<v<1$ but only one such that $\mu>0$. Hence, we find a unique solution and one may note that Eq.~\eqref{pointwise limit} is identical to Eq.~\eqref{transeqmp}.

In order to further compare the two cases, we can rewrite the hypertor correction to inertial motion by using the double-angle formulae for $\sin$ and $\sinh$,
\begin{align}
    \F_{\tc{h}}^{\text{corr}}(E)&~\sim~\frac{a^2}{4\pi\gamma\Omega}\frac{\ee^{-2\tfrac{|E|\mu}{\gamma\Omega}}}{\rho^2\sinh(2\mu)-\alpha\sin(2\alpha\mu)}\,,\\
    ~=~\frac{a^2}{8\pi\gamma\Omega}&\frac{\ee^{-2\tfrac{|E|\mu}{\gamma\Omega}}}{\rho^2\sinh(\mu)\cosh(\mu)-\alpha\sin(\alpha\mu)\cos(\alpha\mu)}\,.
\end{align} Since the solution to~\eqref{transeqm} is bounded as $a\rightarrow0$, we also have that $a\mu\rightarrow0$ in this limit; hence, we can use a small angle approximation, $\alpha\sin(\alpha\mu)\cos(\alpha\mu)\approx\alpha^2\mu$. This approximation and the defining equation~\eqref{pointwise limit} lead to $\alpha\sin(\alpha\mu)\cos(\alpha\mu)\approx\alpha^2R\Omega\sinh(\mu)$. This remark along with the definition $\alpha = a/\Omega$ allow the final factorization of the response rate as
\begin{equation}
    \F_{\tc{h}}^{\text{corr}}(E)~\sim~\frac{1}{8\pi\gamma R}\frac{\ee^{-2\tfrac{|E|\mu}{\gamma\Omega}}}{\sinh\mu(R\Omega\cosh\mu-1)}\,.
\end{equation} This is simply the response rate for a detector in circular motion in a $3+1$ thermal bath~\eqref{eqn: large E rate circ} with $\mu_+\mapsto\mu$. Additionally, the effective temperatures agree to leading order owing to the fact that $\mu\rightarrow\mu_+$ as $a\rightarrow0$. Hence, in the limit of small accelerations/small initial thermal state temperatures, we reconcile the circular motion in a $3+1$ thermal bath with ambient temperature $T<T_{\text{crit}}$\,.

\vspace{-0mm}

\subsubsection{Limit of small circular velocity}\label{subsubsec vel}

We consider now the limit of small velocities within the large-gap regime.  Considering first the circular motion in a thermal bath, one may estimate Eq.~\eqref{transeqT} by noting that as $v\rightarrow0$, the nonzero root to the equation $x=\sinh(vx)$ grows. Using this fact, the exponentially growing part of the hyperbolic sine function is the dominant contribution and one can solve for $T_{\text{crit}}$, giving
\begin{equation}\label{largegapsmallvT}
    T_{\text{crit}}~\sim~-\frac{v}{4RW_{-1}(-\tfrac{v}{2})}\,,
\end{equation}where $W_{-1}(x)$ is the lower branch of the Lambert $W$ function~\cite{database}. $W_{-1}(x)$ diverges to $-\infty$ as $x\rightarrow0^-$, hence $T_{\text{crit}}\rightarrow0^+$ in this limit. As such, we need only consider the behaviour of $\mu_-$ for a general $T$.

A graphical argument shows that the first nonzero root of Eq.~\eqref{transeqmm} tends to zero in the limit $v\rightarrow0$ and one may employ a small-argument expansion. We assume that the root is of the form $\mu_-=a_1v+a_2v^2+a_3v^3+\mathcal{O}(v^4)$. Under this ansatz, one finds
\begin{equation}
    \mu_-~=~\frac{1}{2RT}(v-v^2+v^3)+\mathcal{O}(v^4)\,.
\end{equation} For completion, we note that similar arguments used to derive \eqref{largegapsmallvT} show that
\begin{equation}\label{muplus}
    \mu_+~\sim~-W_{-1}(-\tfrac{v}{2})\,.
\end{equation}
By considering the large-gap detailed balance effective temperature \eqref{templargegap} and~\eqref{muplus}, one finds that the temperature tends to zero, for $T<T_{\text{crit}}$, as one would expect. However, for $T>T_{\text{crit}}$, one finds
\begin{align}\label{tempthermalbathlargegap}
    T_\tc{tb}&~=~\frac{\gamma\Omega RT}{v(1-v+v^2)}+\mathcal{O}(v^3)\,,\\
    &~=~T(1+v+\tfrac{1}{2}v^2)+\mathcal{O}(v^3)\,.\label{circularlargeEsmallv}
\end{align}

We turn our attention now to the hypertor motion and to finding an expansion for the root of Eq.~\eqref{transeqm}. Let $z=\pi-\tfrac{Ra}{v}\mu$. Equation~\eqref{transeqm} then reduces to
\begin{equation}
    \sin(z)~=~Ra\sinh(\frac{v}{Ra}(\pi-z))\,,
\end{equation} which we may expand in small arguments. Using the ansatz $z=a_1v+a_2v^2+\mathcal{O}(v^3)$, one finds that $a_1=-a_2=\pi$ and we can write down an expansion for the root,
\begin{equation}
    \mu~=~\frac{v}{Ra\pi}(1-v+v^2)+\mathcal{O}(v^4)\,.
\end{equation} Combining this with the large-gap detailed balance effective temperature \eqref{largegaptemphyper}, one finds
\begin{align}
    T_\tc{H}&~=~\frac{\gamma\Omega R a\pi}{v(1-v+v^2)}+\mathcal{O}(v^3)\,,\\
    &~=~\frac{a}{2\pi}(1+v+\tfrac{1}{2}v^2)+\mathcal{O}(v^3)\,.\label{hypertorlargeEsmallv}
\end{align}The leading-order contribution can be identified as the Unruh temperature $T_U$. One may reconcile the effective temperature of the circular motion in a thermal bath \eqref{circularlargeEsmallv} and the effective temperature experienced by the detector in hypertor motion \eqref{hypertorlargeEsmallv} via the Unruh temperature $T=a/2\pi$. Remarkably, the two agree to the first three orders.

    \begin{figure*}[t]
\begin{subfigure}{.5\textwidth}
  \centering
  \includegraphics[width=8.6cm]{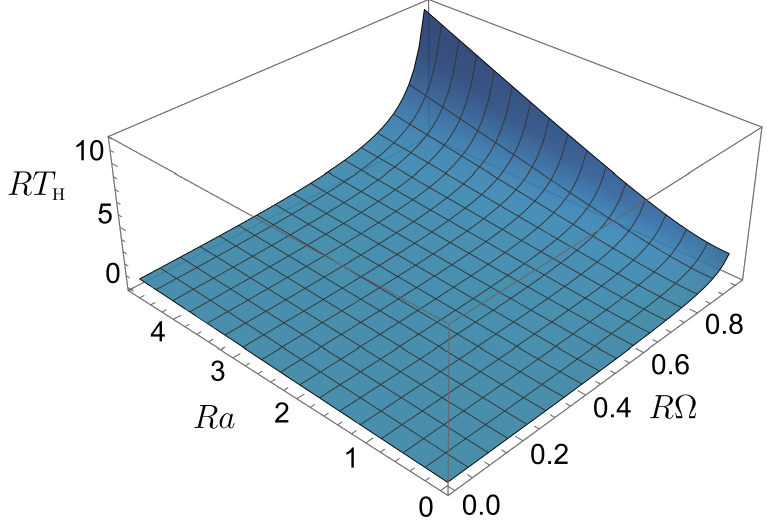}
    \caption{}
    \label{fig:largeE}
\end{subfigure}%
\begin{subfigure}{.5\textwidth}
  \centering
  \includegraphics[width=8.6cm]{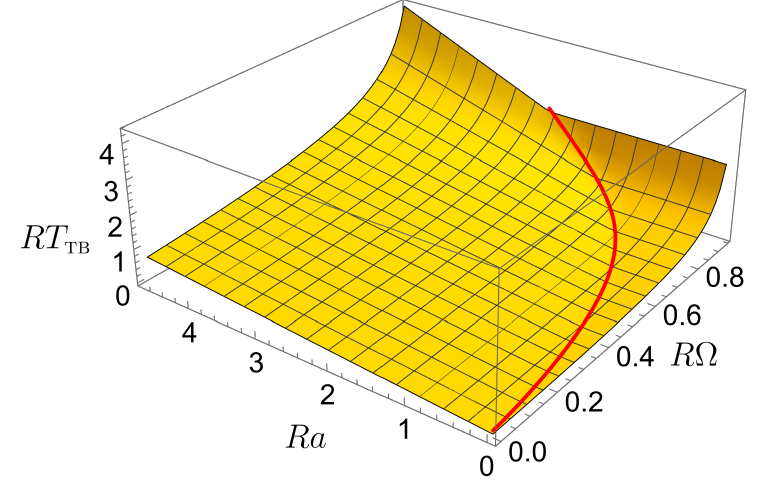}
    \caption{}
    \label{fig:jorma}
\end{subfigure}
\caption{\justifying Effective temperatures experienced by UDW detectors in a) hypertor motion and b) circular motion in a $3+1$ thermal bath in the large-gap limit as functions of dimensionless variables $Ra$ and $R\Omega$. \textbf{a)} Effective temperature experienced by UDW detector in hypertor motion interacting with the Minkowski vacuum. The plotting range is $0\leq Ra\leq4.8$, $0\leq R\Omega\leq 0.91$. \textbf{b)} Effective temperature experienced by the UDW detector in circular motion interacting with a scalar field initially prepared in a thermal state. The plotting range is $0\leq Ra\leq4.8$, $0\leq R\Omega\leq 0.91$. A red line is shown at parameter values satisfying $T=T_{\text{crit}}$ where $T_{\text{crit}}=a_{\text{crit}}/2\pi$.}
\label{fig:Einfinity}
\end{figure*}

\subsubsection{Numerical results}

In Fig.~\ref{fig:Einfinity}, we plot the effective temperatures for the two motions as functions of the dimensionless variables $Ra$ and $R\Omega$: in Fig.~\ref{fig:largeE}, we plot the effective temperature for the hypertor motion; in Fig.~\ref{fig:jorma} we plot the effective temperature for circular motion in a $3+1$ thermal bath at temperature $T=a/2\pi$. The red line in Fig.~\ref{fig:jorma} corresponds to the line of critical temperature~\eqref{transeqT}, across which the leading-order behaviour of the large-gap limit changes. All plots are adimensionalized with respect to the natural length scale $R$.

\section{Conclusions}\label{sec:conclusions}
Motivated by the duality between the experience of a static observer in a thermal bath and a linearly accelerated observer, we have analyzed the effect of composing circular motion with these two situations. In particular, we considered a UDW detector undergoing hypertor motion (circular motion in the plane orthogonal to linear acceleration), interacting with a quantum scalar field in the Minkowski vacuum and compared it with a UDW detector undergoing circular motion, interacting with a quantum scalar field prepared in a thermal state. We worked in the limit of weak interaction and long interaction time and neglected the detector's backaction on the field.

We quantified the experience of the detector in terms of the detector's transition probability, transition rate, and the effective temperature, which related the excitation and de-excitation rates.

We found that, in general, this duality is broken over much of the parameter space by the addition of the circular motion. Specifically, we found analytic expansions for the transition rate and effective temperature of the two motions in three regimes: small energy gap, small orbital radius, and large energy gap. In all three regimes, we found that we could reconcile the two motions when the circular motion parameters were small, or if the acceleration parameter/initial thermal state temperature was small. On the other hand, the two situations can be drastically different. In particular, in the regimes of small energy gap and small radius, the effective temperature experienced by the detector in hypertor motion is always higher than that experienced by a detector undergoing circular motion in a $3+1$ thermal bath at temperature $T=a/2\pi$.

In the limit of small orbital radii for the circular motion, we found that the orbital frequency of the circular motion plays a key role in distinguishing between the two motions. In this regime, a detector in hypertor motion with acceleration parameter $a$ or in circular motion in a thermal bath of temperature $T=a/2\pi$ would only be able to distinguish between the two cases if it were to probe below this resonant frequency, $|E|<|\Omega|$. This relates to previous results of~\cite{bunnyCircular}, where similar resonance effects were discussed. 

We remark that our results on the effective temperature appear to bear no relation to the Tolman scaling, which states that the local temperature in a thermal equilibrium state is proportional to $1/\sqrt{-g_{00}}$ in the coordinates adapted to the notion of stationarity with respect to which the temperature is defined~\cite{TolmanPrelim,TolmanEhrenfest,TolmanBook}. For a corotating detector in a rotating thermal ensemble, the detector's response does scale by the Tolman factor, an example being a corotating detector in the Hartle-Hawking state on the rotating Ba\~nados-Teitelboim-Zanelli black hole, considered in~\cite{bhDetectorsBTZ}. In our system, however, the rotation is just in the detector but not in the ensemble, and there is no reason to expect the response to be related to the Tolman factor. 

We also recall that our analysis operated within linear perturbation theory in the coupling between the detector and the field, and to this order the detector's response requires no renormalisation. We are not aware of a relation between the energy dependence in our results and the energy dependence in the running coupling constants that appear in the renormalisation of scattering amplitudes in perturbative quantum field theory.

In summary, while a particle detector interacting with a massless scalar field in $3+1$ Minkowski spacetime would be unable to distinguish between a thermal bath of temperature $T=a/2\pi$ and uniform acceleration $a$, we found that circular motion breaks this duality. In particular, any small circularity added to these two situations enables a particle detector to tell the difference between a thermal bath and linear acceleration temperature across most of the parameter space.

\vspace{-0.00pt}

\section{Acknowledgements}
We thank the anonymous referee for helpful comments. TRP acknowledges support from the Natural Sciences and Engineering Research Council of Canada (NSERC) via the Vanier Canada Graduate Scholarship. The work of JL was supported by United Kingdom Research and Innovation Science and Technology Facilities Council [grant number ST/S002227/1]. Research at Perimeter Institute is supported in part by the Government of Canada through the Department of Innovation, Science and Industry Canada and by the Province of Ontario through the Ministry of Colleges and Universities. Perimeter Institute and the University of Waterloo are situated on the Haldimand Tract, land that was promised to the Haudenosaunee of the Six Nations of the Grand River, and is within the territory of the Neutral, Anishinaabe, and Haudenosaunee people. For the purpose of open access, the authors have applied a CC BY public copyright licence to any Author Accepted Manuscript version arising.
\\\\
\textbf{Note after publication:} We thank Bibhas Ranjan Majhi for bringing to our attention the work in C. Chowdhury, S. Das, S. Dalui and B. R. Majhi, Phys. Rev. D \textbf{99}, 045021 (2019), in which a similar question was addressed.

\appendix
\onecolumngrid
\section{Transition rates and probabilities}\label{app:rates}

In this appendix we discuss the relationship between the transition rate and the transition probability of a UDW detector interacting with a scalar field, using the UDW model described in Section~\ref{sec:UDW}. Specifically, we are interested in comparing the following three definitions of the transition rate found in the literature,
\begin{equation}\label{F forms}
    \F_1(E) ~=~ \int_\RR \dd\tau\, \ee^{- \ii E \tau} W(\tau)\,, \quad\quad \F_2(E) ~=~ \lim_{T\to \infty} \frac{P(E)}{\lambda^2 T}\,, \quad\quad \F_3(E) ~=~ \frac{1}{\lambda^2}\lim_{T\to \infty}\dv{P(E)}{T}\,,
\end{equation}
where $W(\tau)$ is the Wightman function pulled back to a stationary trajectory~\eqref{stationary wightman}, $P(E)$ is the transition probability~\eqref{eq:PE}, and $T$ is a time parameter that controls the interaction time.

We work with the interaction Hamiltonian of Eq.~\eqref{eq:HI} with switching function $\chi(\tau)$, which we assume can be written as
\begin{equation}
    \chi(\tau) ~=~ \beta(\tau/T)\,,
\end{equation}
where $\beta(u)$ is a real function determining the shape of the switching and the parameter $T$ determines the interaction time. We, furthermore, assume that $\beta(0)=1$ so that in the limit of long interaction times $T\rightarrow\infty$, we recover the infinite time switching function $\chi(\tau)=1$.

Though we consider the model of Section~\ref{sec:UDW}, the results of this appendix are valid for a more general class of quantum field theories and spacetimes. We, however, focus our attention to the case of a detector coupled to a quantum scalar field in an arbitrary spacetime. The key assumption for our results is that the state of the field is stationary with respect to time translations along the detector trajectory, so that the Wightman function is also stationary~\eqref{stationary wightman}. Under these assumptions, the transition probability is written as
\begin{equation}\label{excitation prob}
    P(E) ~=~ \lambda^2 \int_{\RR^2 } \dd \tau \dd \tau'\, \chi(\tau) \chi(\tau')\ee^{-\ii E (\tau - \tau')} W(\tau - \tau')\,.
\end{equation}
We write the pullback of the Wightman function in terms of its Fourier transform as
\begin{align}
    W(\tau - \tau') ~=~ \frac{1}{2\pi}\int_\RR \dd \omega\,  \ee^{\ii \omega (\tau - \tau')}\widehat{W}(\omega)\,.
\end{align}

One should note that due to the singular behavior of the Wightman function at $\tau=0$, the Fourier transform should be understood in a distributional sense. We then rewrite the transition probability~\eqref{excitation prob} as
\begin{align}
    P(E) &~=~ \frac{\lambda^2}{2\pi} \int_\RR \dd \omega\, \widehat{W}(\omega) \left(\int_\RR\dd \tau \, \ee^{- \ii (E - \omega) \tau}\chi(\tau)\right) \left(\int_\RR\dd \tau'\,\ee^{\ii (E - \omega) \tau'}\chi(\tau')\right)\,,\\
    &~=~ \frac{\lambda^2}{2\pi} \int_\RR \dd \omega\, \widehat{W}(\omega) |\widehat{\chi}(E - \omega)|^2\,,\label{prob chi}
\end{align}
where $\widehat{\chi}(\omega)$ is the Fourier transform of $\chi(\tau)$. This may be written as
\begin{equation}
    \widehat{\chi}(\omega) ~=~ \int_\RR \dd \tau\, \ee^{-\ii \omega \tau}\chi(\tau) ~=~ T \int_\RR \dd u\, \ee^{- \ii \omega T u}\beta(u) ~=~ T \widehat{\beta}(\omega T)\,.
\end{equation}
Using this form of $\widehat{\chi}$, we perform the change of integration variable $v=(E-\omega)T$ in~\eqref{prob chi}, such that
\begin{equation}\label{P(E) beta}
    P(E) ~=~ \frac{\lambda^2}{2\pi} T\int_\RR \dd v\, \widehat{W}(E - v/T)|\widehat{\beta}(v)|^2.
\end{equation}
The transition rate $\F_2(E)$~\eqref{F forms} can then be recast as
\begin{equation}
    \F(E) ~=~ \lim_{T\to \infty} \frac{P(E)}{\lambda^2 T} ~=~ \widehat{W}(E)\times \frac{1}{2\pi} \int_\RR \dd v |\widehat{\beta}(v)|^2\,. 
\end{equation}

On the other hand, a second common notion of \textit{transition rate} is $\F_3(E)$~\eqref{F forms}, where a derivative is taken with respect to the time parameter $T$. This can be directly computed using~\eqref{P(E) beta} as
\begin{equation}
    \dv{P(E)}{T} ~=~ \frac{\lambda^2}{2\pi} \int_\RR \dd v\, \widehat{W}(E - v/T)|\widehat{\beta}(v)|^2+\frac{1}{T}\frac{\lambda^2}{2\pi} \int_\RR \dd v\, \widehat{W}'(E - v/T)|\widehat{\beta}(v)|^2\,.
\end{equation}
Under the assumption that the derivative of $\widehat{W}(\omega)$ is {sufficiently regular, as is the case for the Wightman functions considered in this paper}, we take the limit $T\rightarrow\infty$ and recover
\begin{equation}
    \F_3(E) ~=~ \frac{1}{\lambda^2}\lim_{T\to\infty}\dv{P(E)}{T} ~=~ \widehat{W}(E) \times \frac{1}{2\pi} \int_\RR \dd v\, |\widehat{\beta}(v)|^2 ~=~ \F_2(E)\,,
\end{equation}
demonstrating the equivalence of the definitions $\F_2(E)$ and $\F_3(E)$ in~\eqref{F forms}.

Finally, we note that the definition of $\F_1(E)$~\eqref{F forms} is given by $\F_1(E)=\widehat{W}(E)$. In order to reconcile all three definitions, the switching function must be normalized such that 
\begin{equation}\label{beta condition}
    \int_\RR \dd v\, |\widehat{\beta}(v)|^2 ~=~ 2\pi\,.
\end{equation}
In this case, all three definitions of the transition rate agree. This, hence, justifies our choice of switching function discussed in Section~\ref{sec:UDW}, which both satisfies this condition~\eqref{beta condition} and $\lim_{T\to\infty}\chi(\tau;T)=1$.

\bibliography{references}

\end{document}